\documentclass[fleqn,usenatbib]{mnras}

\usepackage{newtxtext,newtxmath}

\usepackage[T1]{fontenc}
\usepackage{ae,aecompl}

\usepackage{graphicx}	
\usepackage{amsmath}	
\usepackage{amssymb}	
\usepackage{multicol}   
\usepackage{threeparttable}
\usepackage{enumitem}

\newcommand{\lSect}[1]{{\label{sec:#1}}}
\newcommand{\lFig}[1]{{\label{fig:#1}}}

\newcommand{\lTab}[1]{{\label{tab:#1}}}
\newcommand{\FIGFF}[2]{{\ref{fig:#2}{#1}}}

\newcommand{\FIG}[2]{{Fig.~\FIGFF{#1}{#2}}}
\newcommand{\Fig}[1]{{\FIG{}{#1}}}

\newcommand{\Sectff}[1]{{\ref{sec:#1}}}
\newcommand{\Sect}[1]{{\S\Sectff{#1}}}

\newcommand{\Msun}{\ensuremath{\mathrm{M}_\odot}}

\newcommand{\Zsun}{\ensuremath{\mathrm{Z}_\odot}}
\newcommand{\Tab}[1]{{Table \ref{tab:#1}}}
\newcommand{\cp}{\ensuremath{\mathrm{\xi}_{2.5}}}
\newcommand{\KEPLER}{\ensuremath{\mathrm{\texttt{KEPLER}}}}
\newcommand{\MESA}{\ensuremath{\mathrm{\texttt{MESA}}}}

\title[BPS Explosion Landscape]{Towards a Realistic Explosion Landscape for Binary Population Synthesis}

\author[Patton \& Sukhbold]{
Rachel A. Patton $^{1}$\thanks{E-mail: patton.502@osu.edu}
and Tuguldur Sukhbold$^{1,2}\thanks{NASA Hubble Fellow}$
\\
$^{1}$Department of Astronomy, The Ohio State University, 140 West 18th Ave, Columbus, OH 43210, USA\\
$^{2}$Center for Cosmology and AstroParticle Physics, The Ohio State University,
191 West Woodruff Avenue, Columbus, OH 43210\\
}

\date{Accepted XXX. Received YYY; in original form ZZZ}

\pubyear{2020}

\begin{document}
\label{firstpage}
\pagerange{\pageref{firstpage}--\pageref{lastpage}}
\maketitle

\begin{abstract}
A crucial ingredient in population synthesis studies involving massive stars is the determination of whether they explode or implode in the end. While the final fate of a massive star is sensitive to its core structure at the onset of collapse, the existing binary population synthesis studies do not reach core-collapse. Instead, they employ simple prescriptions to infer their final fates without knowing the presupernova core structure. We explore a potential solution to this problem by treating the carbon-oxygen (CO) core independently from the rest of the star. Using the implicit hydrodynamics code \KEPLER, we have computed an extensive grid of 3496 CO-core models from a diverse range of initial conditions, each evolved from carbon ignition until core-collapse. The final core structure, and thus the explodability, varies non-monotonically and depends sensitively on both the mass and initial composition of the CO-core. Although bare CO-cores are not perfect substitutes for cores embedded in massive stars, our models compare well both with \MESA~and full hydrogenic and helium star calculations. Our results can be used to infer the presupernova core structures from population synthesis estimates of CO-core properties, thus to determine the final outcomes based on the results of modern neutrino-driven explosion simulations. A sample application is presented for a population of Type-IIb supernova progenitors. All of our models are available at \url{https://doi.org/10.5281/zenodo.3785377}.
\end{abstract}




\section{Introduction}
\lSect{intro}
Massive stars tend to live in binary systems, with 50-70\%, of O-type stars exchanging mass with their companion and a third of mass-transfer binaries merging over the lifetime of the primary star \citep[e.g.,][]{San12,San13}. Binary interactions can drastically alter the evolution of a massive star, its ultimate demise, properties of the transient, and the compact remnant it produces. To understand how true populations of massive stars evolve and die, we must consider binary interactions.

There is a rich literature of binary population synthesis (BPS) studies aimed at doing just that. Some recent examples include studies of the compact object mass distribution \citep[e.g.,][]{Spe15}, of the progenitors of long-duration gamma ray bursts \citep{Chr19}, of binary merger and mass transfer rates \citep{deM14}, of compact object mergers \citep{Bel16}, of supernova rates \citep[e.g.,][]{Sra19,Zap19}, and of the ejection of binary companions \citep[e.g.,][]{Ren19}. These types of investigations play a critical role in interpreting various modern observations, including those from transient surveys, gravitational wave radiation, and proper motions. 

\begin{figure*}
\centering
    \includegraphics[width=1.0\textwidth]{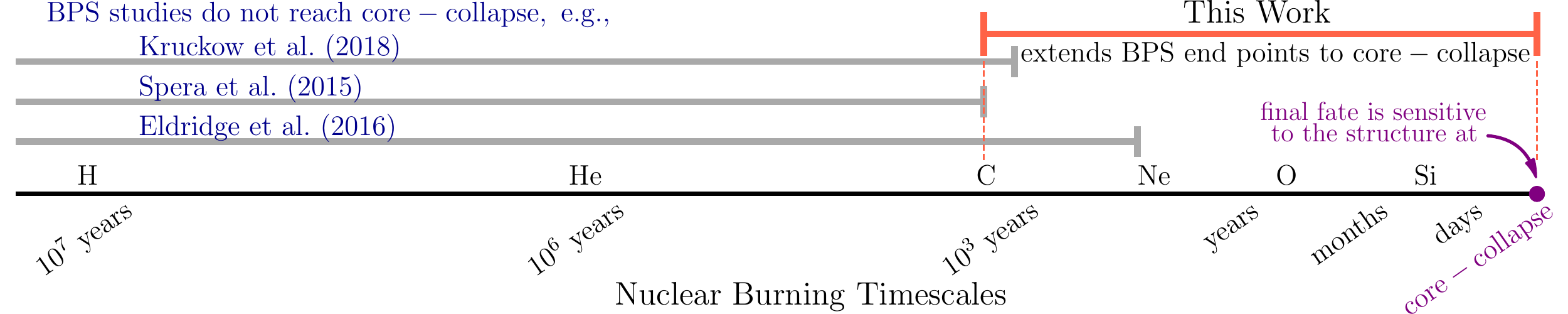}
    \caption{A comparison of the evolutionary stages covered by BPS studies and our bare CO-core models. Although the final fates and the properties of collapse are sensitive to the core structure at the time of collapse (highlighted in purple), BPS studies follow stellar evolution only until carbon or neon ignition in the center (gray bars). Falling short of core-collapse, BPS studies infer the final outcomes through simple prescriptions that are based on the stellar mass and the embedded He- or CO-core masses, which do not cleanly correlate with the final core structure. This work aims to bridge the evolutionary gap, shown in red, between the cutoff of BPS studies and core-collapse, and allows mapping of final outcomes determined from the presupernova core structure back to core properties at carbon ignition.}
    \lFig{comp}
\end{figure*}

Carefully following the complete evolution of both stars in a binary system, from zero-age main sequence (ZAMS) until core-collapse, including their often-complicated interactions, can be a challenging problem for a single binary system, let alone for an entire population. The BPS community employs various innovative ways to circumvent these difficulties, for example, by not evolving the stars until core-collapse, and often by not directly simulating their evolution. The so-called ``rapid'' BPS studies use a semi-analytical approach to approximate stellar evolution through the asymptotic giant branch based on fits from \citet{Hur00,Hur02}, modified to include binary interactions \citep[see, e.g.,][]{Bel02a,Bel02b,DeD04,Izz04,Izz06,Izz09,Gia18a,Gia18b,Bre19}. Some groups actively follow the evolution using stellar evolution codes, but they cut off at carbon or neon ignition to avoid and the advanced stages of evolution \citep[e.g.,][]{Sie13}. Still others interpolate over tables of single star models, modified with binary interaction prescriptions \citep{Eld08,Eld17,Spe15,Spe17,Spe19,Kru18}.

Failing to reach core-collapse is a common problem for BPS studies, especially those investigating the final demise of binary massive stars. It has been known for some time that not only the properties of the explosion, but the very fate of the star, whether it explodes or not, is closely tied to the presupernova structure of the core of the massive star \citep{Bur95,OCo11}. This final structure, moments before collapse, is largely set by the advanced stages of evolution in the core, from carbon ignition until the iron-core collapse \citep[e.g.,][]{Woo02}. However, BPS studies never follow this crucial and final part of the star's life taking place inside the carbon-oxygen (CO) core during the final few thousand years (see \Fig{comp}). Evolution typically stops at carbon ignition or, at the latest, neon ignition. Therefore, nearly all BPS studies indirectly infer the final fates and the properties of the explosion \emph{without} calculating the presupernova core structure.

Instead, BPS calculations (and many studies of populations of single stars as well) attempt to determine which stars die in a supernova, as well as the masses of their compact remnants, by using the ZAMS mass, the mass of the star at the evolutionary cutoff, and the embedded He- or CO-core masses individually or in some combination. For instance, \citet{Hur00} introduced a method to infer the remnant types and their respective masses largely from the CO-core mass, while \citet{Eld16} used the prescription by \citet{Heg03} to determine the type of remnant based on the He-core mass, and applied simple energy constraints to estimate their masses. One commonly employed approach, especially in rapid BPS studies \citep[see e.g.,][]{Vig18}, is the ``delayed'' prescription by \citet{Fry12}, which uses the final stellar mass inferred for the presupernova stage in combination with the embedded CO-core mass. More recently, \citet{Zap19} assumed all stars that form a core more massive than the one embedded in a star with an initial mass of 20 \Msun\ would not produce a supernova. These methods are not uniform and the results can be difficult to interpret, especially when final fates are not based on actual core structure.

Using ZAMS mass, final stellar mass, or CO-core mass to infer the final properties is also generally too simplistic. For instance, it is well known that the evolution of the CO-core not only depends on its mass but also depends strongly on its initial composition. Two stars that form CO-cores of the same mass but with different initial core compositions can end up with very different final structures, and thus final fates. Moreover, these methods do not capture the non-monotonic variation of the presupernova core structure with mass that emerges from the interplay between convective burning episodes of carbon and oxygen \citep{Suk14,Suk18,Suk20}. One of the main implications is that there is no single threshold mass that cleanly delineates explosion vs. implosion, i.e. no CO-core or ZAMS mass where all lighter stars are likely to die in a supernova and all heavier stars likely to implode. Calibrated neutrino-driven explosion surveys of single stars \citep[e.g.,][]{Pej15,Suk16} and of stripped cores in close binary systems \citep{Ert19} have demonstrated that the non-monotonically varying final core structures significantly affect the resulting remnant masses, nucleosynthesis, and light curves.

We explore one potential solution to this problem by carrying out the evolution of the stellar core starting from the point where BPS studies typically stop all the way until they experience core-collapse. Due to extremely short nuclear burning timescales, we can approximately treat the evolution of the CO-core independently from the rest of the star. Because of its fast pace, the ``accelerated'' evolution of the CO-core is also fairly immune to major uncertainties of stellar evolution. Here we create a large suite of constant mass CO-cores covering a wide range of possible combinations of starting mass and composition, each evolved from the ignition of carbon in the center until the presupernova stage. In reality, of course, the CO-core mass is not truly constant, the ``frozen-envelope'' approximation is not entirely valid, and certain rare types of binary interactions may directly alter the evolution of the core. Nevertheless, this approach provides a simple way to connect the endpoints of most stellar models in BPS studies to specific presupernova core structures.

Just knowing the final core structure for each star in a BPS simulation does not automatically produce more accurate results for final fates, but it does open the door to explore more physics-based methods. There have been a number of studies in the past decade, with varying complexity, attempting to connect the presupernova core structure with the final fate of the star and the properties of the supernova explosion. The simplest, and perhaps the most commonly employed, approach is the compactness parameter \citep{OCo11}, which attempts to peg the structure outside of the iron-core into a single parameter, $\xi_{2.5}$, the inverse of the radius enclosing innermost 2.5 \Msun. A more careful approach to determine the final fates was suggested by \citet{Ert16}, using both $M_4$, the mass encompassed at the location where the entropy per baryon equals four, and $\mu_4$, the radial gradient of mass at that location. There is also a more detailed semi-analytical method by \citet{Mue16} that attempts to capture some of the key ingredients of the core-collapse supernova problem, including the heating, core accretion, shock revival, and neutrino flux, in order to predict the final fates and masses of the compact remnants. These, and other approaches, have been extensively studied in the context of neutrino-driven explosion scenario \citep[e.g.,][]{Ugl12,Suk16,Pej15,Ert19,Ebi19,Mab19}, and our models allow these recent insights to be reflected in BPS calculations.

In this paper, we fill in the evolutionary gap in BPS simulations between carbon ignition and core-collapse, in order to explore a more accurate and comprehensive landscape of final fates based on the actual presupernova core structure. We evolve the CO-cores of massive stars at the time of carbon ignition from a dense grid of initial masses and compositions through core collapse using \KEPLER, an implicit hydrodynamics code. Using the structure of the core immediately preceding collapse, we calculate the properties of each model, including the subset of the parameters described above. We discuss these models in detail in \Sect{num}. We present a table of ``explodability'' as a function of core mass and starting composition in \Sect{expl}. To verify these results, we run a subset of the core models using the open source stellar evolution code \MESA\ and, in \Sect{mesa}, provide comparison with the \KEPLER\ results. As an additional test, in \Sect{comp}, we compare the presupernova properties from the bare CO-cores to those from sets of full hydrogenic stars and helium star models depicting stripped cores in close binary systems. We describe the implementation and application of these results in \Sect{impl}. Finally, in \Sect{concl} we summarize our results and briefly outline future plans.


\section{CO Core Evolution with \KEPLER}
\lSect{num}

\subsection{Motivations for using CO Cores}
\lSect{coc}
Simulating cores independently from the rest of the stars for computational studies is not new. Half a century ago He- and O-cores were being used to investigate the late-stage evolution of massive stars \citep[e.g.,][]{Arn72a,Arn72b}. Initially, cores offered a less computationally expensive means to approximate and probe the intricate physics leading up to core-collapse \citep{Nom88,Bar90}. While the computational expense is no longer as much of an issue as before \citep[see, e.g.,][and references therein]{Yoo10,Suk14,Woo19}, the separate treatment of cores remain useful approach because the evolution of the core is not subject to the complete range of uncertainties affecting full stars (see below).

Treating the stellar core in isolation from the rest of the star relies on the ``frozen-envelope'' approximation, that the cores are effectively decoupled from their envelopes in the late stages of evolution. With the ignition of carbon, when the central temperature exceeds roughly $5\times10^8$ K, the stellar core begins to rapidly cool by strong neutrino emission. As a consequence of the temperature sensitivity of neutrino losses, and the need to reach higher temperatures to burn heavier fuels, the evolution of the CO-core is drastically accelerated \citep{Woo02}. From the ignition of carbon, it takes less than a few thousand years to reach iron-core collapse. The lifetime of the CO-core, essentially defined by the carbon burning timescale, is much shorter than the Kelvin-Helmholtz timescale of the envelope, especially if it is puffy and extended. The stellar envelope hardly has time to respond to the rapid changes in the core, and therefore, the evolution of the envelope is largely disconnected from that of the embedded CO-core.

One would expect core-envelope decoupled stars to evolve quiescently from the formation of the core until collapse. Observations of red supergiant supernova progenitors do show that this is the case \citep[e.g.,][]{Joh18}, however, not all massive stars go out so quietly. There is clear evidence to suggest that some uncertain fraction of supernova progenitors experience outbursts, some violent, in the millennium preceding core-collapse \citep[e.g.,][]{Pas07,Fra13,Mau13,Koc17}. Theoretically, it is well established that very massive stars experience late stage instability due to pair-production \citep[e.g.,][]{Woo17}. At the lowest masses, the stars may experience instabilities due to silicon-flashes \citep{Woo15}, and at intermediate masses, the convective burning episodes in the core during the late stages of evolution may be able to transport significant amount of energy to the surface by exciting gravity waves \citep{Shi14,Ful17}. In the absence of such instabilities, however, the CO-cores are effectively decoupled from their envelopes.

Once the CO-core forms, its largely independent evolution is characterized by two key parameters, (1) the mass of the CO-core, and (2) its initial uniform composition of carbon and oxygen \citep[e.g.,][]{Suk20}. These two parameters are set by the preceding main sequence and core helium burning evolution, which can be dramatically affected by various uncertain processes, such as mass loss \citep[e.g.,][]{Ren17}, rotation \citep[e.g.,][]{Lim18}, complex binary interactions \citep[e.g.,][]{Woo19,Men17}, and key uncertain reactions such as $^{12}\mathrm{C}(\alpha,\gamma)^{16}\mathrm{O}$ and $3\alpha$ \citep[e.g.,][]{Tur07}. However, once the CO-core forms and carbon ignites, these processes are either irrelevant or the CO-core evolves too rapidly for them to have a significant effect. Therefore, by just focusing on CO-cores spanning a wide range of initial conditions, agnostic of the processes producing those conditions, we can then largely ignore some of the major uncertainties of stellar evolution.

There are several caveats to our approach. First and foremost, a bare CO-core is not a perfect substitute for a core embedded in a full star. Unlike embedded helium cores, there is no steep pressure gradient spanning several orders of magnitude at the boundary of the CO-core. The mass of the CO-core is not exactly constant, and more importantly, the outer parts of real CO-cores enclose the deepest parts of the overlaying helium burning shell. We discuss these issues in further detail in \Sect{comp}. Secondly, regardless of whether or not the core is embedded in an envelope, its evolution is still sensitive to a limited set of uncertainties. For instance, changes in certain nuclear reaction rates \citep[e.g., $^{12}$C+$^{12}$C,][]{Ben12} and in the treatment of mixing at the convective boundaries \citep{Mea06,Mea07} can alter the late-stages of evolution. Finally, we are not sensitive to any binary interaction that occurs after carbon ignition. While the lifetime of a star is dominated by hydrogen and helium burning, certain rare types of Case C mass transfers could lead to very late mergers, a scenario somewhat similar to the one proposed for the progenitor of SN 1987A \citep{Pod07}.

Nevertheless, it is the CO-core that is at the heart of late-stage evolution. Most physical processes (binary interaction, rotation, mass loss, nuclear reaction rates, etc.), however uncertain, are largely manifested in changing the starting mass and composition of the CO-core, which in turn sets its unique evolutionary path until core-collapse. Therefore, by evolving a large grid of CO-cores until core-collapse from varying starting points, we can attempt to effectively link the final presupernova structure surrounding the iron-core back to the mass and composition of the core at carbon ignition. In conjunction with modern BPS calculations, these results can be used to link the presupernova structures to whichever complicated evolutionary pathways produced these CO-core initial conditions.

\subsection{\KEPLER\ models}
\lSect{kepler}

Using the implicit hydrodynamics code \KEPLER\ \citep{Wea78}, we run a comprehensive suite of 3496 non-rotating CO-cores, each evolved from the ignition of carbon until iron-core collapse. Carbon is ignited roughly when the central temperature exceeds $\sim 5\times10^8$ K, and the onset of core-collapse is defined as the point in evolution where the infall velocity exceeds 1000 km s$^{-1}$ anywhere in the core. The input physics configurations are largely the same as in \citet{Suk18}. A small 19-isotope network is employed until oxygen depletion in the center, and a 121-isotope quasi-equilibrium network is used afterwards. We impute a constant boundary pressure of $10^{12}$ dyne cm$^{-2}$ applied throughout the evolution to simulate the pressure from the overlaying matter. We assume that the CO-core mass remains constant and its initial composition is a pure, uniform mixture of carbon and oxygen only. These are reasonable simplifications, since the ultimate goal is to use these results in order to reveal underlying trends in populations of stars, rather than trying to accurately capture the evolution of each individual model.

The CO-core mass, $M_{\rm CO}$, is varied between 2.5 and 10 \Msun\ in increments of 0.1 \Msun. The lower bound is larger than the minimum CO-core mass that can evolve until iron-core formation, estimated to be around $\sim$1.4 \Msun, corresponding to ZAMS mass of approximately $\sim$9 \Msun\ \citep[e.g.,][]{Woo15}. We avoid the lightest iron-core producing CO-cores (< 2.5 \Msun) since they are more degenerate and can be computationally challenging. However most or all of them likely will result in successful supernova explosions, based on simulations of the neutrino driven explosions of light massive stars \citep[e.g.,][]{Mel15,Suk16,Ert19,Bur20}. The upper limit of our survey, set to 10 \Msun, is well below the limit where the pulsational pair-instability effects become relevant, which is estimated to be around 30 \Msun\ by \citet{Woo19}. In between 10 and 30 \Msun, the CO-cores will be very difficult to explode, and they will all likely collapse into black holes, in the neutrino-driven scenario \citep[e.g.,][]{Ert19,Suk16}. As we will show in \Sect{expl}, the range between 2.5 and 10 \Msun\ captures the region of parameter space where the final outcomes are most likely to vary due to the non-monotonically varying final core structures.

The initial uniform composition is defined by a carbon mass fraction, $X_{\rm C}$,  which varies from 0.05 to 0.5 in increments of 0.01. As we assume a pure mixture of only carbon and oxygen, the mass fraction of oxygen is $1-X_{\rm C}$. In stars, the initial composition of the CO-core is almost always oxygen-rich, with $X_{\rm C}<0.5$, and the mass fractions are strongly influenced by the competition between $3\alpha$ and $^{12}$C$(\alpha,\gamma)^{16}$O during the preceding core helium burning phase. The rate of the three-body reaction $3\alpha$ drops due to the higher entropy in more massive cores, and thus typically, $X_{\rm C}$ is highest in smaller mass cores and decreases with increasing $M_{\rm CO}$.

Populations of single stars for a given set of input configurations occupy well defined, narrow bands in the $X_{\rm C}-M_{\rm CO}$ plane (illustrated and discussed further in \Sect{comp}). For instance, the red supergiant models of \citet{Suk18} smoothly vary from  $M_{\rm CO}=2.1$ \Msun\ and $X_{\rm C}=0.25$ to $M_{\rm CO}=8.2$ \Msun\ and $X_{\rm C}=0.18$ (their Fig. 7). Helium stars evolved with mass loss from \citet{Woo19} range from $M_{\rm CO}=2.3$ \Msun\ and $X_{\rm C}=0.34$ to $M_{\rm CO}=12.2$ \Msun\ and $X_{\rm C}=0.22$ (their Figs. 7 and 14). The same parameters range from $M_{\rm CO}=2.0$ \Msun\ and $X_{\rm C}=0.37$ to $M_{\rm CO}= 33$ \Msun\ and $X_{\rm C}=0.09$ in the survey on rotating massive star models with enhanced mass loss by \citet[][their Fig.19]{Lim18}. These variations are all covered by our grid. Although the extreme cases of high mass and high $X_{\rm C}$ and low mass and low $X_{\rm C}$ may not exist in nature, we have attempted to cover a wide parameter space because the outcomes from  BPS calculations may occupy a much more diverse region of parameter space than any population of single stars.

\begin{table*}
    \centering
    \caption{Sample entries from the table of $M_\mathrm{Fe}$, $M_4$, and \cp\ all evaluated at the onset of collapse, as a function of CO-core mass (columns) and carbon mass fraction (rows) at the time of carbon ignition. The full tables and the final structure and composition for each of the 3496 CO-core models are available online.}
    \begin{tabular}{ccccccccc}
    \hline
         {} & {} & {} & {} & $M_\mathrm{Fe}\:(\Msun)$ & {} & {} & {} & {}\\
    \hline
         $_{X_{\rm C}} \textbackslash ^{M_{\rm CO}}$ & 2.5$\:\Msun$ & 3.5$\:\Msun$ & 4.5$\:\Msun$ & 5.5$\:\Msun$ & 6.5$\:\Msun$ & 7.5$\:\Msun$ & 8.5$\:\Msun$ & 9.5$\:\Msun$ \\
    \hline 
         0.05 & 1.38 & 1.58 & 1.78 & 1.82 & 1.50 & 1.48 & 1.65 & 1.62\\
         0.15 & 1.41 & 1.52 & 1.72 & 1.83 & 1.78 & 1.57 & 1.80 & 1.77\\
         0.25 & 1.47 & 1.56 & 1.44 & 1.55 & 1.39 & 1.67 & 1.85 & 1.54\\
         0.35 & 1.36 & 1.33 & 1.37 & 1.43 & 1.43 & 1.44 & 1.61 & 1.66\\
         0.45 & 1.41 & 1.35 & 1.36 & 1.32 & 1.45 & 1.64 & 1.65 & 1.76\\
   \hline
        {} & {} & {} & {} & $M_4\:(\Msun)$ & {} & {} & {} & {}\\
   \hline
        $_{X_{\rm C}} \textbackslash ^{M_{\rm CO}}$ & 2.5$\:\Msun$ & 3.5$\:\Msun$ & 4.5$\:\Msun$ & 5.5$\:\Msun$ & 6.5$\:\Msun$ & 7.5$\:\Msun$ & 8.5$\:\Msun$ & 9.5$\:\Msun$ \\
   \hline
        0.05 & 1.47 & 1.85 & 2.28 & 2.24 & 1.85 & 1.63 & 1.77 & 1.90\\
        0.15 & 1.62 & 1.74 & 2.16 & 2.21 & 2.08 & 2.06 & 2.21 & 2.23\\
        0.25 & 1.61 & 1.68 & 1.71 & 1.63 & 1.57 & 2.00 & 2.38 & 1.66\\
        0.35 & 1.40 & 1.47 & 1.40 & 1.55 & 1.43 & 1.44 & 1.94 & 1.76\\
        0.45 & 1.46 & 1.44 & 1.50 & 1.29 & 1.46 & 1.73 & 1.77 & 1.84\\
   \hline
        {} & {} & {} & {} & $\xi_{2.5}$ & {} & {} & {} & {}\\
   \hline
        $_{X_{\rm C}} \textbackslash ^{M_{\rm CO}}$ & 2.5$\:\Msun$ & 3.5$\:\Msun$ & 4.5$\:\Msun$ & 5.5$\:\Msun$ & 6.5$\:\Msun$ & 7.5$\:\Msun$ & 8.5$\:\Msun$ & 9.5$\:\Msun$ \\
   \hline
        0.05 & 0.06 & 0.30 & 0.51 & 0.59 & 0.38 & 0.22 & 0.26 & 0.31\\
        0.15 & 0.08 & 0.20 & 0.48 & 0.47 & 0.35 & 0.30 & 0.45 & 0.50\\
        0.25 & 0.08 & 0.20 & 0.19 & 0.21 & 0.13 & 0.38 & 0.53 & 0.27\\
        0.35 & 0.05 & 0.07 & 0.06 & 0.09 & 0.06 & 0.06 & 0.35 & 0.26\\
        0.45 & 0.04 & 0.06 & 0.04 & 0.04 & 0.14 & 0.25 & 0.33 & 0.37\\
   \hline     
        
    \end{tabular}
    \lTab{results}
\end{table*}


\section{Explodability of CO-Cores}
\lSect{expl}

Two final outcomes dominate the fates of massive stars that experience iron-core collapse: (1) a successful supernova leaving behind a neutron star, and (2) an implosion event without a bright transient that promptly forms a stellar mass black hole. Though there are number of other possibilities, with or without iron-core collapse, they are rare and not expected to significantly influence the overall trends in populations of massive stars. Pair-instability supernovae, which leave behind no remnant, happen only at very high mass, and thus are intrinsically rare. Electron-capture supernovae, which may leave behind a neutron star \citep[or sometimes a white dwarf, e.g.,][]{Jon16}, may happen at lower mass, but the relevant mass range is highly uncertain and could be too narrow to be significant \citep{Jon13,Woo15}. Though binary interactions can widen its effective initial mass range, the overall properties of these explosions are not too different from lower energy iron-core collapse supernovae. Additionally, light black holes could be formed through delayed massive fallback in iron-core collapse supernovae, however, recent neutrino-driven explosion surveys indicate that they are infrequent \citep{Ert16,Suk16,Ert19}. Therefore, we only consider the two outcomes that are expected to be the most common: explosions that make neutron stars and prompt implosions that make black holes.

In the most general sense, the existing neutrino-driven explosion surveys indicate that lighter massive stars (corresponding to about $1.4~ \Msun \lesssim M_\mathrm{CO} \lesssim 6~\Msun$) tend to explode, while higher mass stars tend to form black holes. The overall fraction of implosions is about $\sim30\%$ in a sample population of single full stars \citep{Suk16,Suk18}, and of mass losing helium stars that depict stripped cores in close binary systems \citep{Woo19,Woo20}. The two main outcomes are not cleanly separated in mass-space, however. Instead, the results exhibit a complicated explosion landscape \citep[e.g., see Fig.13 of][]{Suk16}, with ``islands'' (in mass-space) of implosions at lower mass, and ``islands'' of explosions at higher mass. These outcomes are closely correlated with the final core structure, which dictates the dynamics of the ensuing collapse and can directly facilitate or impede the launch of an outgoing shock.

The origin of this complexity is that the characteristics of central carbon burning depend sensitively on the initial conditions of the CO-core, and these changes propagate throughout the evolution until collapse to produce drastically different final structures, sometimes even for CO-cores that evolved from similar starting points. For a general background on the advanced stage evolution we refer readers to many existing studies \citep[e.g.,][]{Woo02,Suk14,Suk18,Woo19,Suk20}. Here we only highlight some of the key aspects, and concentrate the discussion on the explodability of CO-cores based on their final structures before the collapse. A detailed analysis on the advanced stage evolution will be presented in a separate study.

\subsection{Core structure at the onset of collapse}
\lSect{struct}

We focus on three, closely related, simple parameters that probe the final core structure: the iron-core mass ($M_{\rm Fe}$), the mass point where the entropy per baryon (in units of the Boltzmann constant) exceeds four going outward ($M_4$), and the compactness parameter evaluated at the location enclosing the innermost 2.5 \Msun\ \citep[$\xi_{2.5}$,][]{OCo11}. For simplicity, we define $M_\mathrm{Fe}$ to be the enclosed mass where the fraction of silicon drops below 1\%, going inward\footnote{The results are nearly identical to the values obtained from alternative definitions, such as those based on infall velocity or the electron mole number profile \citep[e.g.,][]{Heg01,Far16}}. Generally stars that form lighter iron-cores with a sharply declining external density profile explode more easily. The entropy jump sampled by the $M_4$ point is almost always carved out by the last oxygen burning shell, which marks a rapid change in the density profile outside the iron-core, and in successful explosion simulations this point strongly correlates with the baryonic mass of the neutron star. Usually stars are easier to explode when the entropy jump is large and is located deep (smaller $M_4$) in the core. The compactness parameter directly samples the density profile outside the iron-core by simply measuring the inverse of the radius enclosing the innermost few solar masses. The value of this parameter is sensitive to the arbitrarily chosen time and mass points of evaluation, 2.5 \Msun\ and presupernova stage respectively for this study, but the results do not qualitatively change with other choices, i.e. higher values correspond to stars more difficult to blow up and vice versa. All parameters are evaluated at the onset of collapse, when the infall velocity exceeds $1000$ km s$^{-1}$ anywhere in the core.

\begin{figure}
    \centering
    \includegraphics[scale=0.44,trim={0.5cm 0.1cm 1.2cm 1.0cm},clip]{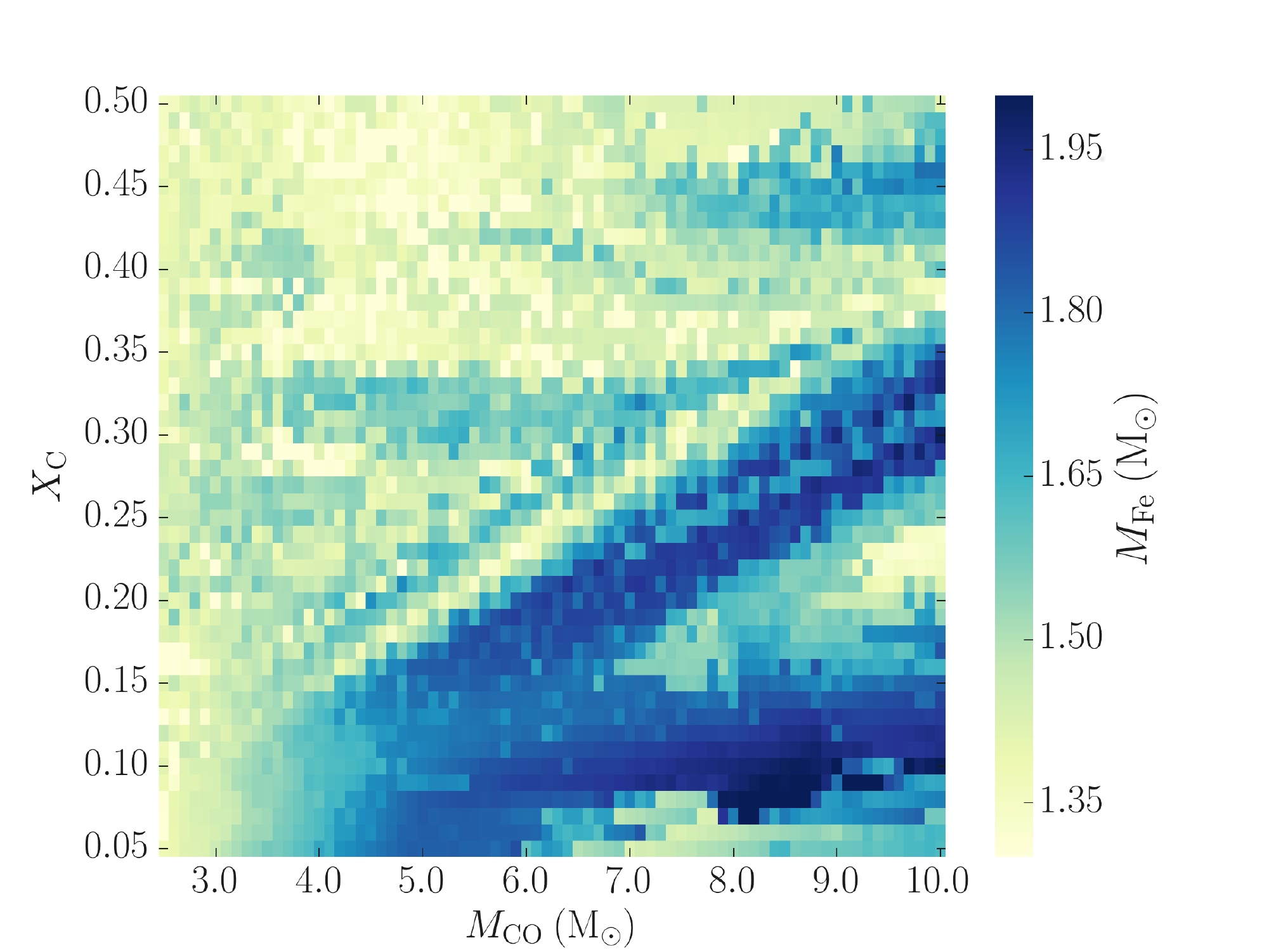}
    \includegraphics[scale=0.44,trim={0.5cm 0.1cm 1.2cm 1.0cm},clip]{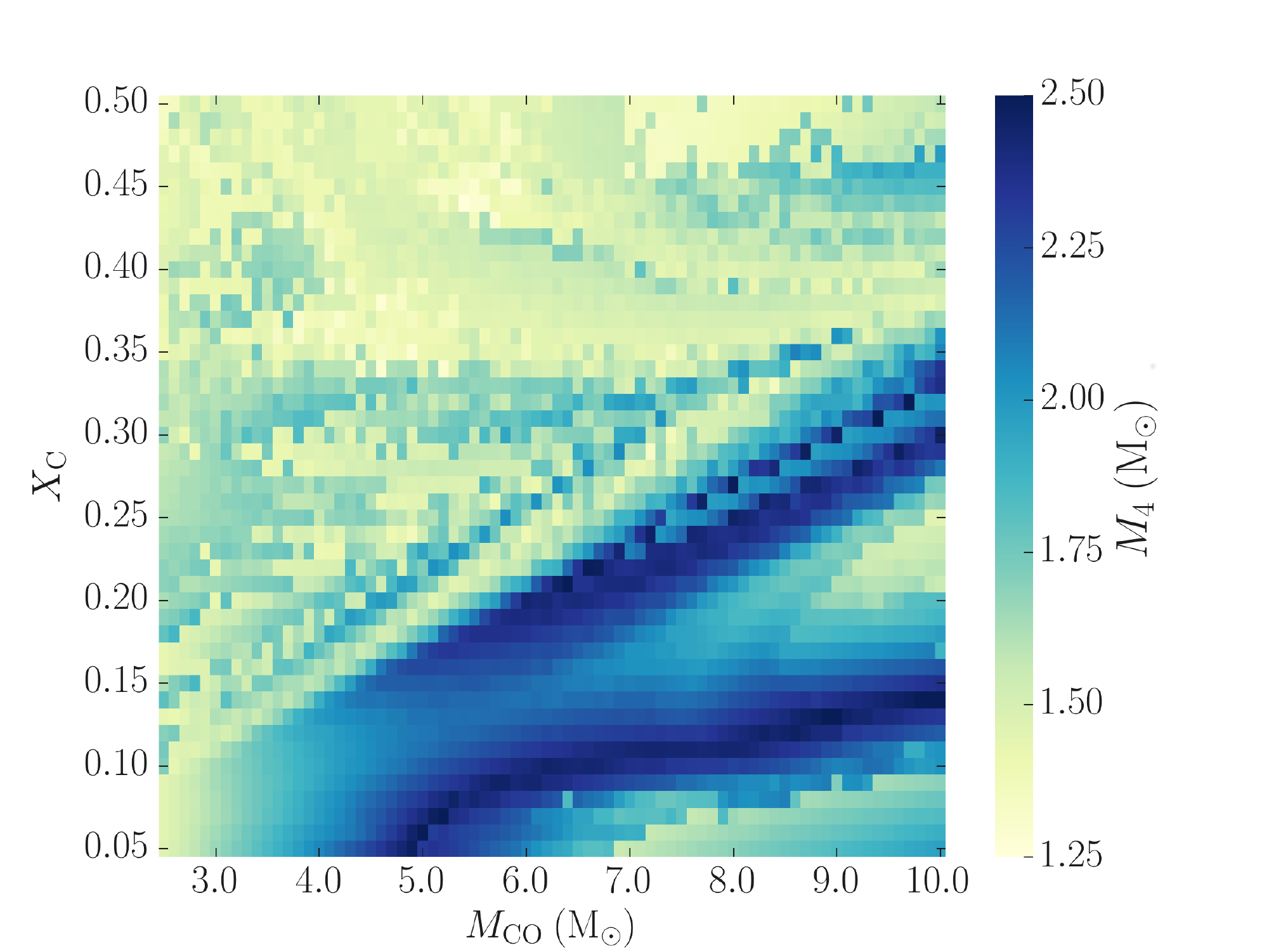}
    \includegraphics[scale=0.44,trim={0.5cm 0.1cm 1.2cm 1.0cm},clip]{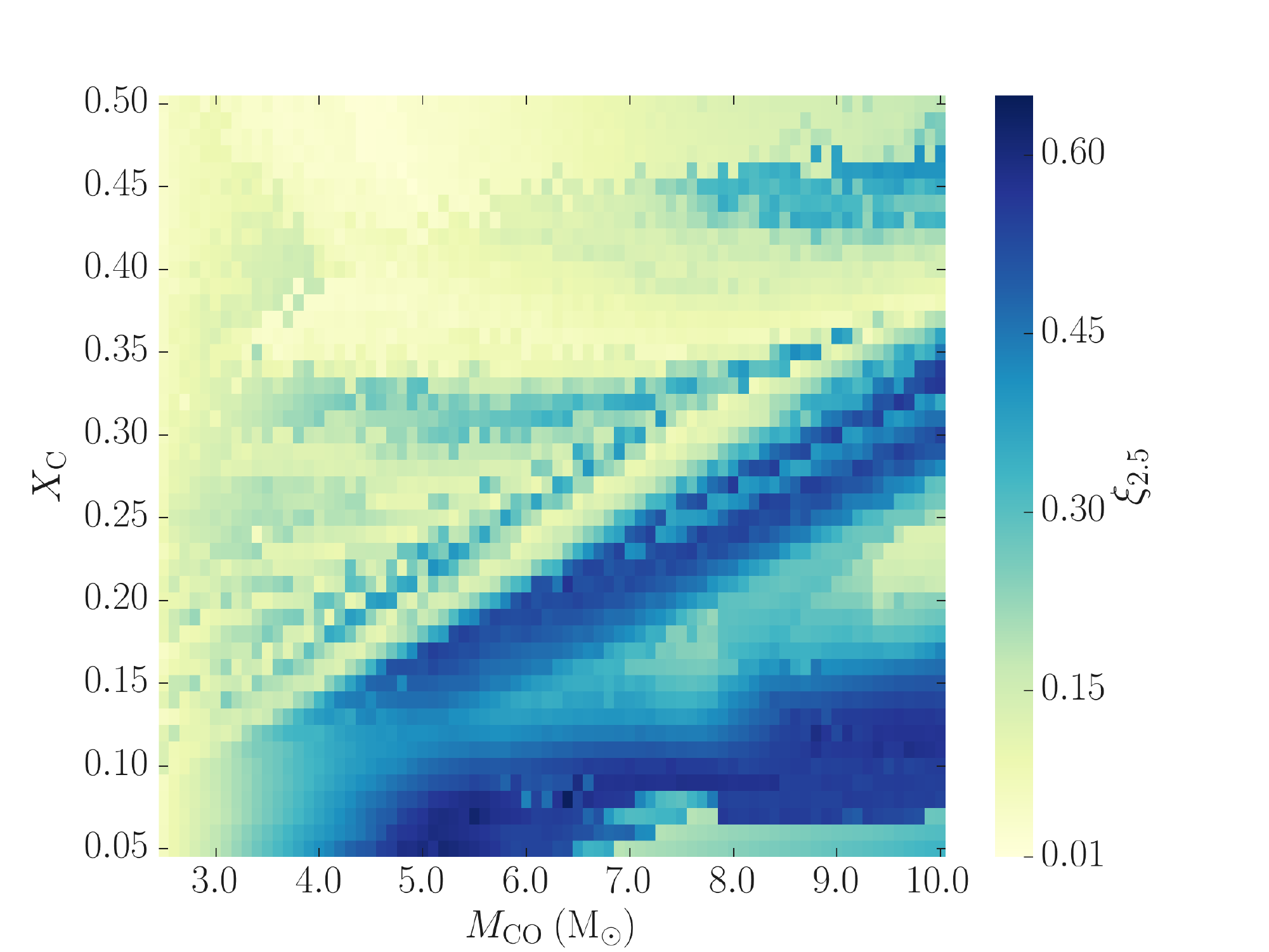}
    \caption{Final core structures from 3496 \KEPLER\ CO-core models with varying mass and initial composition illustrated by their iron-core masses (top), $M_\mathrm{4}$ (middle), and compactness parameter (bottom). The same generic trends in all panels reflect the close correlation between the three parameters; higher values (darker) correspond to cores that are more likely to implode, and lower values (lighter) correspond to cores that are likely to explode. Carbon burns convectively in the center for models above the diagonal line stretching from $M_\mathrm{CO} \approx$ 2.5 \Msun\ and $X_\mathrm{C} \approx$ 0.10 to about $M_\mathrm{CO} =$ 10.0 \Msun\ and $X_\mathrm{C} \approx$ 0.40.}
    \lFig{3par}
\end{figure}

The distributions of these three parameters are illustrated in \Fig{3par}, all plotted with the same color contour. A subset of the corresponding numerical values is listed in \Tab{results}. Over the entire range of models, the iron-core masses range from 1.1 to 2.1 \Msun, $M_4$ ranges from 1.3 to 2.5 \Msun, and $\xi_{2.5}$ is bounded between 0.02 and 0.64. Higher values of all three parameters (darker colors) correspond to CO-cores that are more difficult to blow up, and vice versa. The striking similarity of these panels reflect the close correlation between these parameters. With smaller mass iron-cores, the external density profile falls off steeply, and the overlaying oxygen-burning shell is located deep in the core. The CO-cores that collapse with more massive iron-cores have oxygen burning shells located farther out, and the density outside the iron-core remains high. As has been pointed out in prior studies \citep[][and references therein]{Suk20}, higher values of $\xi_{2.5}$ typically correspond to higher $M_4$, $M_{\rm Fe}$, and lower $\xi_{2.5}$ corresponds to lower $M_4$ and $M_{\rm Fe}$.

These distributions in the $X_\mathrm{C}-M_\mathrm{CO}$ plane are broadly divided into two major regions. At lower initial masses and higher $X_\mathrm{C}$, stars are easier to explode (lower $\xi_{2.5},\ M_4,\ M_{\rm Fe}$) while at higher initial mass and lower $X_\mathrm{C}$ (higher $\xi_{2.5},\ M_4,\ M_{\rm Fe}$), stars are harder to explode. These broad regions are separated by a narrow diagonal line stretching from $M_\mathrm{CO} \approx$ 2.5 \Msun\ and $X_\mathrm{C} \approx$ 0.10 to about $M_\mathrm{CO} =$ 10.0 \Msun\ and $X_\mathrm{C} \approx$ 0.40. This diagonal line roughly tracks the starting mass and composition of the CO-core where the central carbon burning transitions from the convective to the radiative regime. Carbon burns convectively in the center in the models above this line, and it burns as a radiative flame below. To drive a convective episode, the energy losses due to neutrinos needs to be exceeded by the local energy generation rate from carbon burning, which depends on both the initial mass and composition of the CO-core. At low $X_{\rm C}$, the CO-core is essentially an oxygen core. There is little carbon to burn, let alone to drive a convective episode. At higher CO-core masses, higher carbon mass fractions are required to sustain convection during central carbon burning, causing the upward diagonal trend. When carbon burns radiatively in the center, its entropy is higher because it effectively skips the long lasting neutrino-cooling phase. In higher entropy cores, oxygen generally burns in a very massive convective episode, which ultimately forms a massive iron-core with a shallow external density profile \citep[for a detailed discussion see][]{Suk20, Suk14}.

Although this transition is a major inflection point, it is not the only property of central carbon burning that controls the final core structures. The timing and the mass extent of the convective burning episode or the radiative flame also depends on the starting mass and composition of the CO-core. These changes then affect the timing and extent the next shell and core burning episodes and can propagate throughout the evolution to drastically change the final structure at the onset of core-collapse, even in cores that started from nearly identical initial conditions. These effects are illustrated through the models located below the diagonal line that are easier to explode (lower $\xi_{2.5},~M_4,~M_{\rm Fe}$), and models above the line which have higher $\xi_{2.5},~M_4,~M_{\rm Fe}$. Notice those dark ``streaks'' in models with convective central carbon burning, and prominent bright ``valleys'' that cut through the dark models with radiative carbon burning. The main ``valley'' which roughly lies along the line running from $M_{\rm CO}=3,\ X_{\rm C}=0.05$ to $M_{\rm CO}=10,\ X_{\rm C}=0.25$ is caused by a modulation between the carbon-burning shell and the oxygen-burning core. For these conditions, the carbon burns in a shell located just outside the effective Chandrasekhar mass and prevents the development of a massive oxygen-burning core. A lighter oxygen-burning core then results in a lighter iron-core. The smaller ``valley,'' represented by the most massive ($M_{\rm CO}>7$ \Msun) oxygen-rich CO-cores ($X_{\rm C} < 0.1$), is caused by the same effect, but, in this case, the oxygen-burning shell impedes the development of massive silicon-burning core, which again results in a lighter iron-core. The existence and scope of some of these details are sensitive to the adopted convective mixing physics \citep{Suk14,Suk18}.

\begin{figure}
    \centering
    \includegraphics[scale=0.51,trim={0.0cm 0.1cm 1.3cm 1.3cm},clip]{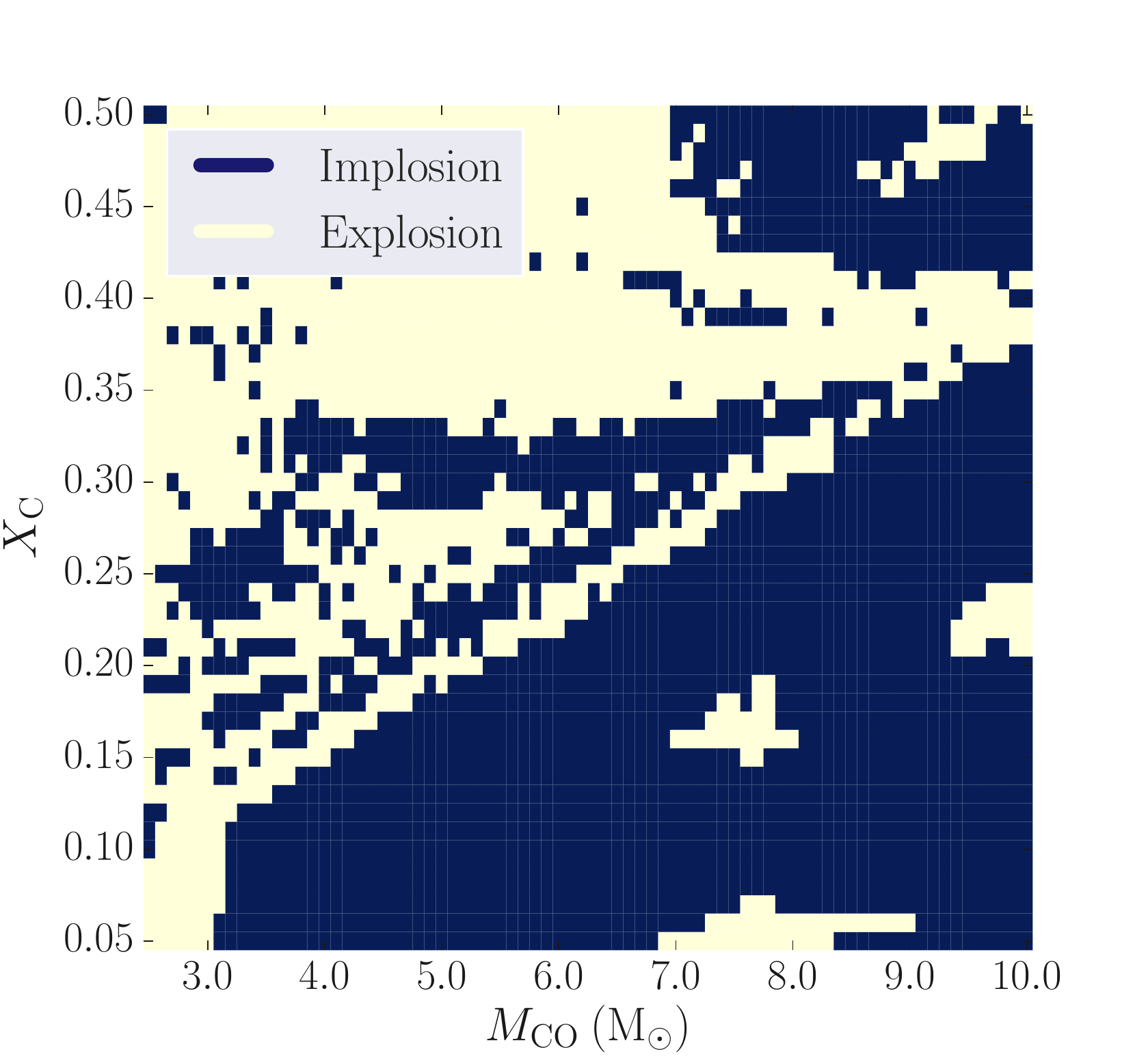}
    \includegraphics[scale=0.42,trim={0.1cm 0.3cm 0.1cm 0.1cm},clip]{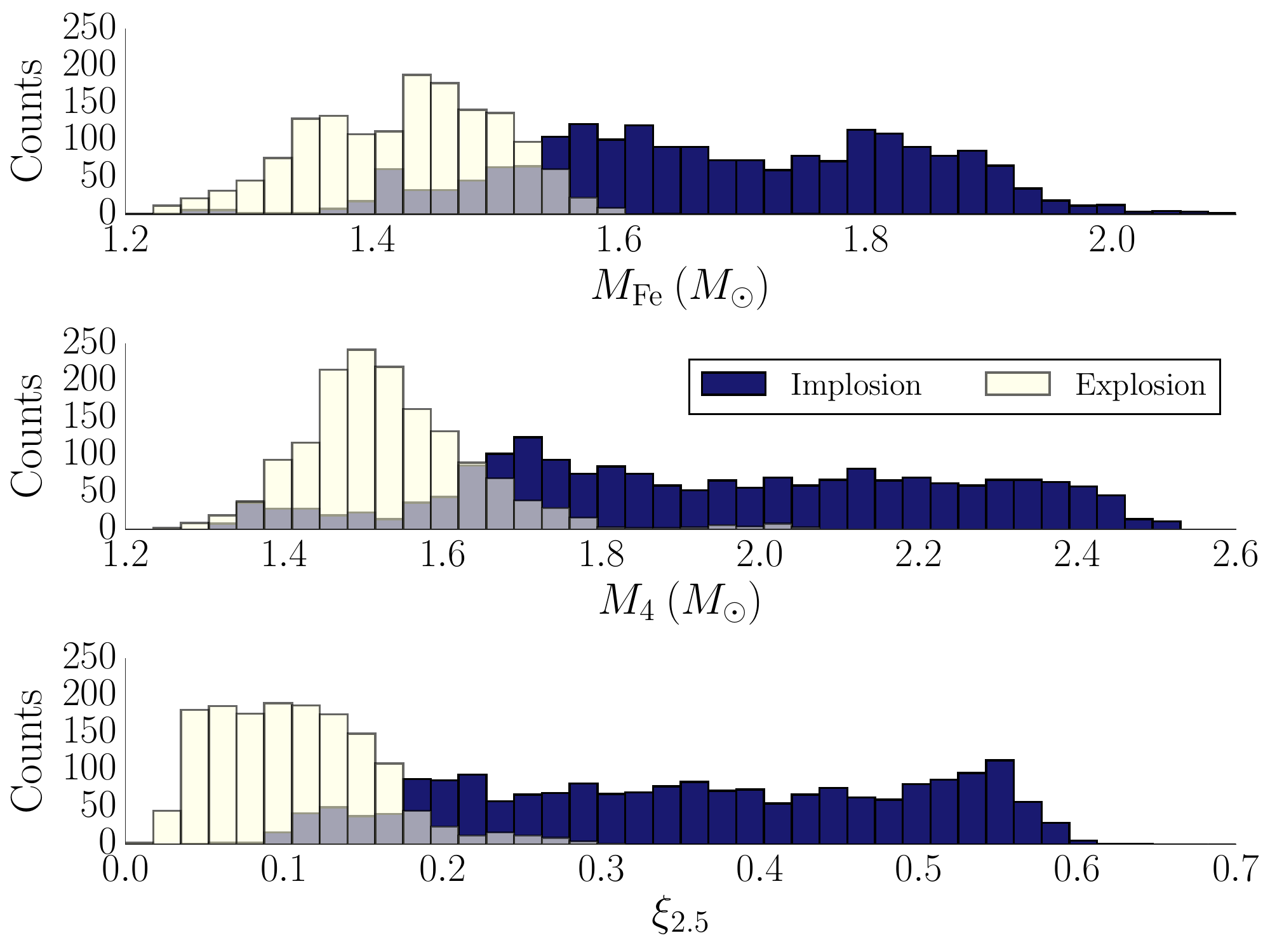}
    \caption{Top: A distribution of the models in the $X_\mathrm{C}$ - $M_\mathrm{CO}$ plane which explode (pale yellow) and implode (dark blue) given a sample explosion criterion from \citet{Ert16,Ert19}. The results closely follow the trends seen in \Fig{3par}, with imploding models tracking higher values of $\xi_{2.5}$, $M_4$, $M_{\rm Fe}$, and vice versa. The final outcomes are closely correlated with final core structure and sensitively depend on both the mass and initial composition of the CO-core. Bottom: Histograms for each parameter evaluated at collapse showing the distribution of values for the models which explode (pale yellow) and implode (dark blue). The crossing points (\Tab{ertl_tab}) of these distributions roughly represent the final structure of models near the central carbon burning transition, and near the critical line of the adopted Ertl criterion.}
    \lFig{ertl}
\end{figure}

\subsection{Assessing the Final Fates: Explosion vs Implosion}
\lSect{ertl}

As our understanding of the neutrino-driven explosion mechanism has not fully converged \citep[e.g., see review by ][]{Jan16}, and given the uncertainties of stellar evolution (\Sect{coc}), currently there is no simple parameter that definitively predicts the final fate of a massive star based on its final core structure. However, we can employ some of the existing neutrino-driven explosion surveys to infer the general trends for populations of massive stars.

As a sample case, here we use the ``Ertl criterion'', developed to reproduce results based on a large number of calibrated one-dimensional explosion simulations of massive stars \citep{Ert16,Ert19}. This method employs a combination of two parameters to probe the final core structure, $M_4$, and the radial gradient of mass at that location, $\mu_4$. The $\mu_4$ parameter measures the amplitude of the entropy jump located by $M_4$, and correlates with the mass accretion rate onto the shock. The explosion criterion is defined as the line $\mu_4 = k_1 M_4 \mu_4 + k_2$, where $k_1$ and $k_2$ are constants. We use the updated calibration for the \texttt{N20} engine with $k_1 = 0.182$ and $k_2 = 0.0608$ \citep{Ert19}. The radial gradient is evaluated within 0.3 \Msun\ of the $M_4$. All cores with $\mu_4$ values above the line form black holes and cores with $\mu_4$ below the line die in a supernova explosion.

The final fates of our models based on the Ertl criterion are shown in \Fig{ertl}, with light boxes representing explosions and dark boxes representing implosions. The models which implode largely trace those with high values of $\xi_{2.5}$, $M_4$, $M_{\rm Fe}$, and vice versa, roughly following the generic trends seen in \Fig{3par}. The diagonal line tracing the transition of the central carbon burning stands out, and we find three distinct ``islands'' of explodability that roughly correspond to the ``valleys'' of \Fig{3par}. Many of these detailed features above and below the carbon burning transition line are not absolutes, and their properties can change depending on the criterion used to determine final fate as well as the input physics of the models. With a weaker engine the ``islands'' may almost disappear, while with a stronger engine they will broaden into valleys, and the dark patches above the diagonal line will diminish. However, nowhere in this distribution can a line be drawn that cleanly separates the two outcomes, which clearly demonstrates that final fates depend on both $M_\mathrm{CO}$ and $X_\mathrm{C}$.

The bottom panel of \Fig{ertl} shows the distributions of the two outcomes for each parameter, with their respective numerical values listed in \Tab{ertl_tab}. With the adopted prescription, no explosions occur for $\xi_{2.5}>0.3$, $M_{\rm Fe}>1.6$ \Msun, and no implosions for $\xi_{2.5}<0.1$, $M_{\rm Fe}<1.35$ \Msun. The $M_4$ values are appreciably more spread out, but generally, there are hardly any explosions above $M_4 > 1.8$ nor implosions below $M_4 < 1.3$ \Msun. The crossing points of these distributions, $\xi_{2.5}\sim0.18$, $M_4\sim1.66$\Msun~, $M_{\rm Fe}\sim1.54$ \Msun, separate \emph{most} of the explosions from \emph{most} of the implosions. They roughly track the final core structures near the central carbon burning transition line and also the critical line separating the two outcomes in the Ertl criterion \citep{Suk20}. Although these values are not too different from the limits and cutoffs discussed in earlier studies \citep[e.g.,][]{Maz82,Woo95,OCo11,Hor14}, we again stress that these should not be taken as absolutes.

Here we demonstrate the evaluation of final fates using a sample criterion for explosion. The full tables containing our numerical results, including $\mu_4$, though it is not listed in \Tab{results}, as well as the structure and composition of each presupernova core are available for download \footnote{\url{http://doi.org/10.5281/zenodo.3785377}}. Readers can take these models and apply their preferred calibrations and criteria for determining final fate. Though the results may change from what is presented here, the bulk trends will stay the same. 

\begin{table}
\centering
\caption{Critical values based on sample explosion criterion}
\begin{threeparttable}
\begin{tabular}{lccc}
\hline
& $M_{\rm Fe}$       & $M_4$              & $\xi_{2.5}$ \\
& $[{\rm M}_{\sun}]$ & $[{\rm M}_{\sun}]$ &             \\
\hline
& \multicolumn{3}{c}{Explosion}\\
\cline{2-4}\vspace{-2mm}\\
min & 1.10 & 1.25 & 0.02 \\
max & 1.74 & 2.10 & 0.30 \\
& \multicolumn{3}{c}{Implosion}\\
\cline{2-4}\vspace{-2mm}\\
min & 1.24 & 1.33 & 0.04 \\
max & 2.10 & 2.52 & 0.64 \\
& \multicolumn{3}{c}{Cross Points}\\
\cline{2-4}\vspace{-2mm}\\
& 1.54 & 1.66 & 0.18 \\
\hline
\lTab{ertl_tab}
\end{tabular}
\begin{tablenotes}
\item Note 'Cross Points' refers to value at which the two distributions cross (see \Fig{ertl}).
\end{tablenotes}
\end{threeparttable}
\end{table}


\section{Validation of Results}
\lSect{test}

In order to confirm our approach, we recreate the \KEPLER\ calculations presented in (\Sect{expl}) with an independent open source stellar evolution code, and also compare our results with full star and helium star models that embed equivalent CO-cores. A good overall qualitative agreement is found in both comparisons. There are differences, however, that point to the limitations of our approach, and highlight that these results should only be used to infer the overarching trends in populations of stars rather than individual models. 

\begin{figure*}
    \centering
    \includegraphics[scale=0.48,trim={0.0cm 0.1cm 0.5cm 1.2cm},clip]{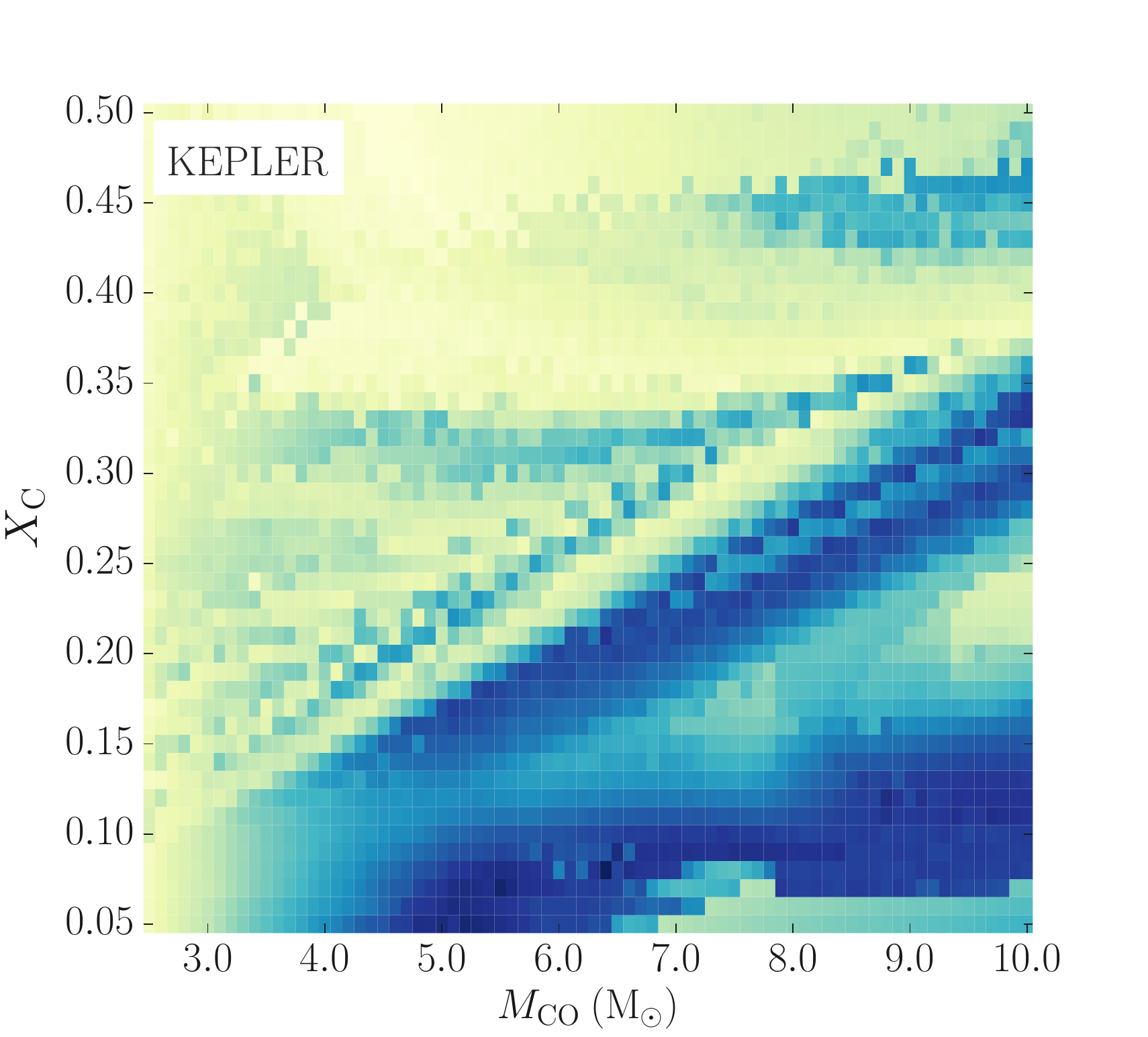}
    \includegraphics[scale=0.50,trim={0.3cm 0.1cm 1.7cm 1.5cm},clip]{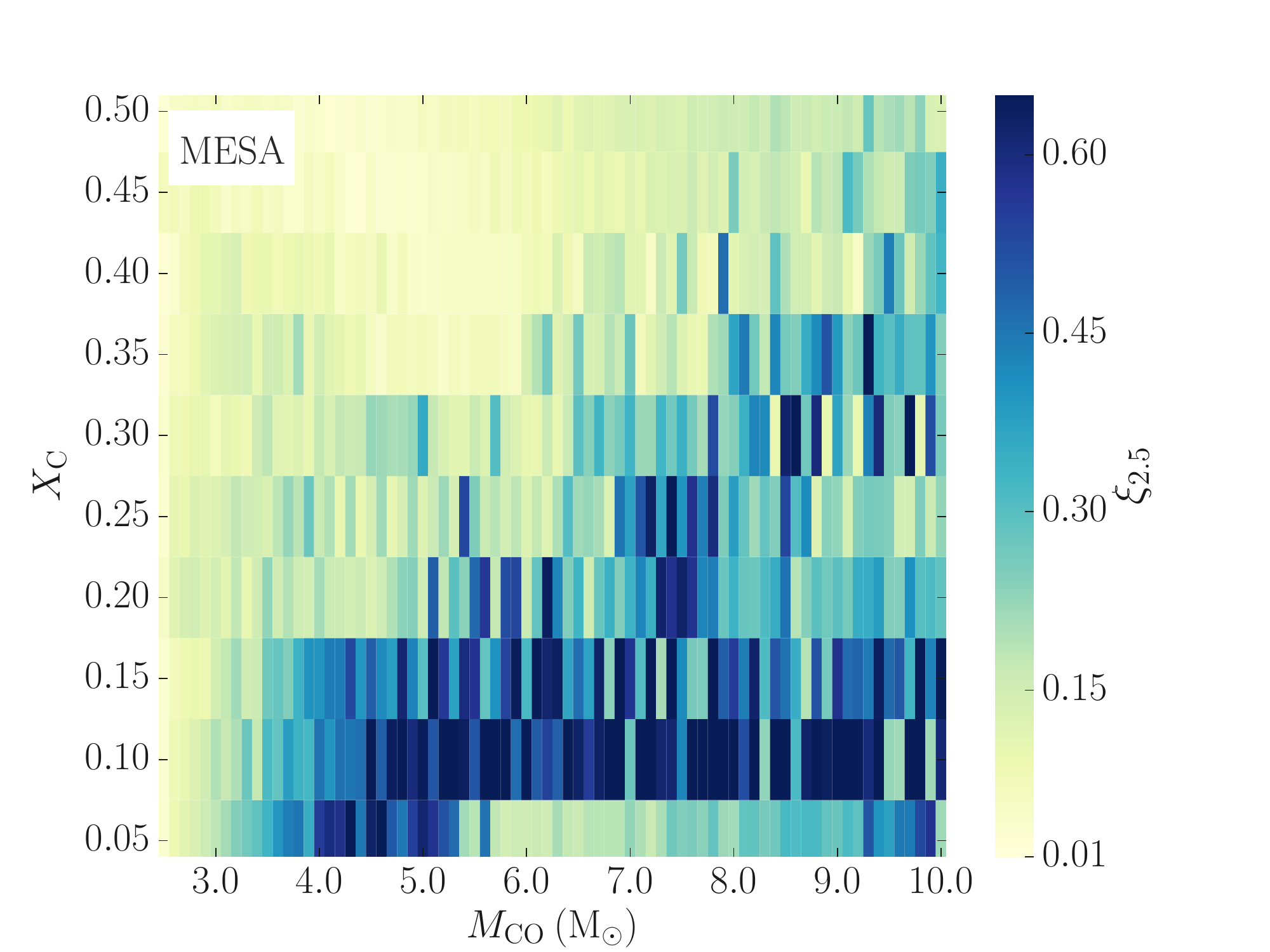}
    \caption{The compactness parameter (\cp) evaluated at core-collapse for our 3496 \KEPLER\ models (left, equivalent to \Fig{3par}) and 760 \MESA\ models (right). The more pixelated appearance of the \MESA\ suite is due to the larger increment in $X_\mathrm{C}$. Similar general trends are evident in both suites, including the upward diagonal line marking the transition of central carbon burning, and ``valleys'' of lower \cp\ beneath this line. A comparison of $M_4$ and $M_\mathrm{Fe}$ show similar behavior and agreement between two codes.}
    \lFig{mesa1}
\end{figure*}

\begin{figure*}
    \centering
    \includegraphics[scale=0.4,trim={4.9cm 0.1cm 5.0cm 1.7cm},clip]{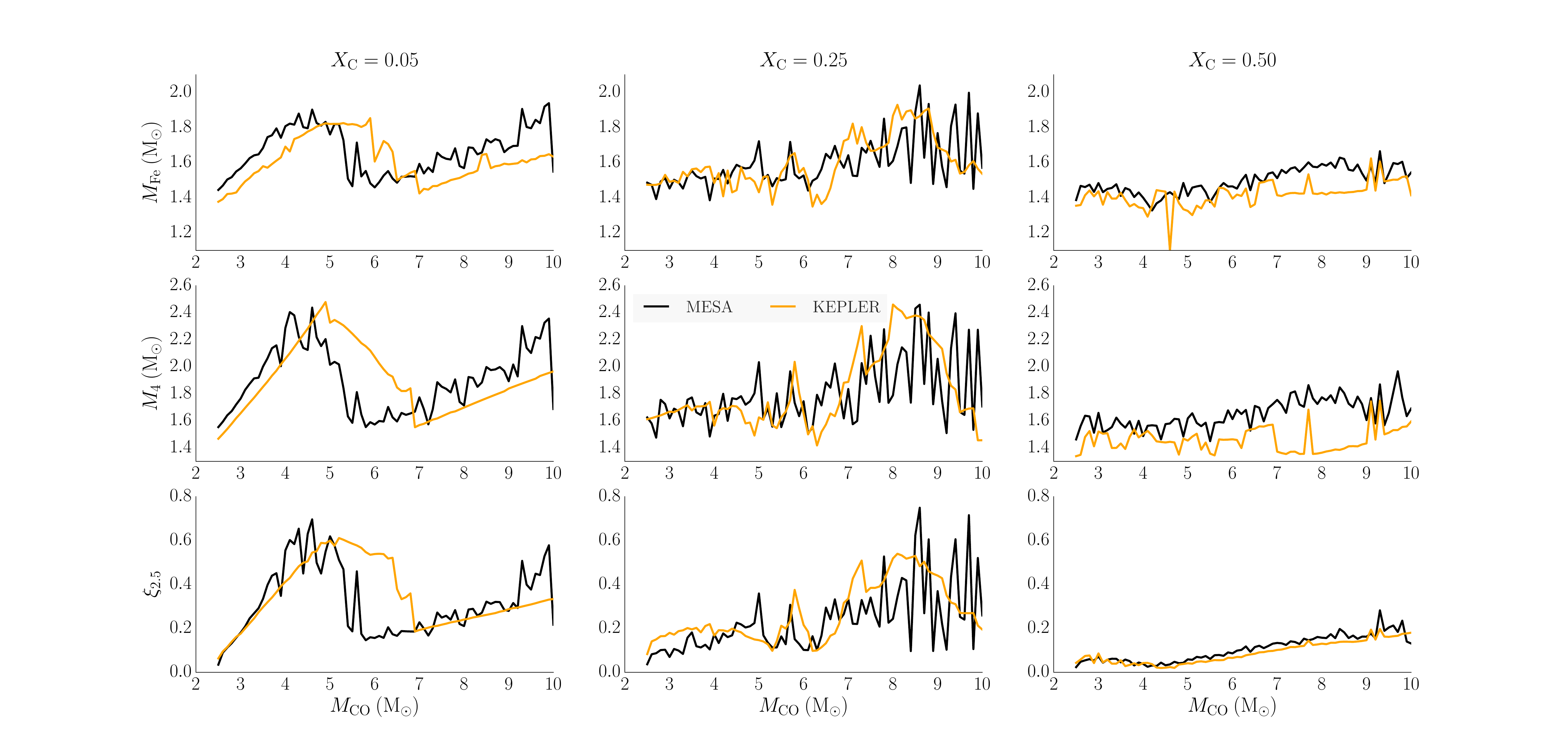}
    \caption{Comparison between \KEPLER\ (orange) and \MESA\ (black) results for all three parameters evaluated at core-collapse (rows) for fixed initial carbon fractions (columns, $X_\mathrm{C} = 0.05, 0.25, 0.50$). Since the CO-core mass increments were identical, there are 76 models per code in each panel. Despite the slight shifts in the structures due to different treatments of convective mixing, and higher oscillations in \MESA\ models due to the use of higher tolerances in the calculations, the general trends are in good overall agreement.}
    \lFig{mesa2}
\end{figure*}

\subsection{Comparison with \MESA}
\lSect{mesa}

We employ version 7624 of the Modules for Experiments in Stellar Astrophysics 1D \citep[\MESA,][]{Pax10,Pax19} to create an equivalent suite of CO-core models. As with the \KEPLER\ grid, the constant CO-core mass ranges between 2.5 and 10 \Msun, and the initial uniform composition ranges between 0.05 and 0.5 in $X_{\rm C}$. Although we maintained the same mass increments (0.1 \Msun), a larger compositional increment was used (0.05 instead of 0.01) for quicker computation. This leads to 760 non-rotating CO-core models all evolved from carbon ignition until core-collapse, which is defined the same way as in \KEPLER\ models. We also employed a small 21-isotope nuclear reaction network. We emphasize that no attempts were made to mimic the results from \KEPLER, though it is possible \citep[e.g., section 4 of ][]{Suk14}. The input configurations, including the convective overshoot mixing setup, are fairly similar to those in recent studies of massive stars \citep[e.g.,][]{Far16}, although the tolerances have been generally relaxed in order to smoothly achieve iron-core formation in large number of many models. The input file (\texttt{inlist}) is included in the online data release \footnotemark[2].

\Fig{mesa1} compares the compactness parameter ($\xi_{2.5}$) evaluated for the \KEPLER\ (same as in \Fig{3par}) and \MESA\ models. The overall good agreement between the two is evident. The \MESA\ models show the same regional trends in this parameter space: a darker region and a lighter region separated by an upward diagonal line, where carbon burns convectively in the center above it, and radiatively beneath. The ``valleys'' beneath the central carbon burning transition line (\Sect{struct}) for higher mass cores ($M_{\rm CO}>5$ \Msun) with lower initial carbon mass fraction ($X_{\rm C}<0.3$) also fall roughly in the same place. These results indicate that the structure of the presupernova cores are similar, despite the different codes and input configurations. Although only the compactness parameter is shown in \Fig{mesa1}, the same good qualitative agreement is also found for the iron-core mass and $M_4$.

A more detailed comparison is presented in \Fig{mesa2}, where all three parameters are compared for different compositions of $X_\mathrm{C} = $ 0.05, 0.25, and 0.50. Since the mass increments were identical, there are 76 models per code in each panel. Though the general trends are in good agreement, the final structures for oxygen-rich CO-cores in \MESA\ (low $X_\mathrm{C}$) move to the ``valley'' (\Sect{struct}) at lower core mass than the \KEPLER\ models. At higher $X_\mathrm{C}$ the $M_4$ values and iron-core masses in the \KEPLER\ models are often lower, indicating a deeper location for the last strong oxygen burning shell. These systematic differences can be largely attributed to the different treatment of convective mixing, which is set to be more efficient in the \MESA\ models \citep[for the sensitivity see][]{Suk14}. Also the \MESA\ solutions generally exhibit larger irregularities and oscillations for all masses and compositions. This is due to the much larger tolerances adopted for the calculations.

\subsection{Comparison with Full Stars and He cores}
\lSect{comp}

To further gauge the validity of our approach, we compare our results to the CO-cores in the suite of single star models from \citet{Suk18} and He star models evolved with mass loss from \citet{Woo19}. The single stars consist of 1500 models with initial masses ranging between 12 and 27 \Msun, all of which retained a significant hydrogen envelope (e.g., died as red supergiants). Their embedded CO-cores marginally grow in mass due to the overlaying helium burning shell. In contrast, the helium stars represent stripped He-cores of massive stars in close binary systems, where the entire hydrogen envelope is removed near the time of helium ignition at the center. The set had 56 models with \emph{initial He-core} masses between 5.00 and 18.75 \Msun. These cores evolve by losing mass in winds, and with their adopted prescription, the embedded CO-cores were exposed before the collapse for He-cores that had initial masses more than about 17 \Msun\ ($M_{\rm CO}\sim7\ \Msun$). Both sets of models were computed with \KEPLER\ and had similar input physics to the models presented in this study.

To make the comparison, we first map these models in the $X_{\rm C}-M_{\rm CO}$ plane by measuring the carbon mass fractions and CO-core masses for each model when the central temperature reaches $5\times10^8$ K (central carbon ignition). The composition is well mixed by the prior convective helium burning core and thus there is no ambiguity in measuring $X_{\rm C}$. However, the mass of the CO-core can be sensitive to how one determines its outer boundary. The boundary of $M_{\rm CO}$ is usually measured at the location where the mass fraction of $^4$He drops below 1\% going inward. This threshold generally works, but it can significantly underestimate the CO-core mass in cases where the $^4$He profile exhibits an extended tail deep into the core. A higher threshold of 20\% or 50\% would typically coincide with the outer boundary of the helium burning shell, but it can also overestimate the $M_{\rm CO}$ in some rare cases. Given these considerations and the simple nature of our models, we chose to measure the CO-core mass as the average of the masses inside the 1\% and 25\% thresholds. 

The results are illustrated in panel (A) of \Fig{full_comp}. Both sets of models occupy a well defined narrow band where the CO-cores begin their evolution with more oxygen (lower $X_{\rm C}$) with increasing mass, due to the dominance of $^{12}$C($\alpha,\gamma$)$^{16}$O over $3\alpha$ during the prior helium burning phase. The single star models span up to about $M_{\rm CO}\sim7.5$ \Msun, while the helium stars reach nearly 9 \Msun. The initial compositions are drastically different, however, due to the receding convective helium burning core in the mass-losing helium star models \citep{Woo19}. Unlike embedded helium cores, which grow in mass, the shrinking convective core leaves behind a gradient of carbon, brings no extra helium into the burning shell, and results in less destruction of carbon. Both sets of models used a rate for $^{12}$C($\alpha,\gamma$)$^{16}$O equivalent to about 1.2 times the rate by \citet{Buc96}.

Based on the location of these models in the $X_{\rm C}-M_{\rm CO}$ plane, we obtain their corresponding final core structure properties by linearly interpolating on our grid of CO-core models. The results comparing the final compactness parameter (outcomes are similar for $M_{\rm Fe}$ and $M_4$) are shown in panels (B) and (C) of \Fig{full_comp}. The agreement with He star set is excellent (panel B), where the integrated values of \cp\ stay low until about $M_{\rm CO}\sim6$ \Msun, after which the CO-cores burn carbon radiatively in the center and form the peak of \cp. Though it is slightly broader and taller, the interpolated values peak at the same mass $M_{\rm CO}\sim7$ \Msun\ and at similar \cp$\sim0.5$. The agreement with full stars (panel C) is good until the embedded cores transition to radiative carbon burning at about $M_{\rm CO}\sim4.5$ \Msun, however, the same transition happens in bare cores at a slightly larger mass. As a result the peak is shifted by about a solar mass, but it is much broader in bare CO-cores. For both of these cases the interpolated values closely track these models especially at lower mass, and over the entire comparison grid it exhibits qualitatively same non-monotonic variation with a distinct peak. 

While this result is highly encouraging, there are some discrepancies which directly point to the limitations of our approach. \Fig{full_comp} shows that the values of \cp~show the biggest differences at higher CO-core masses. At a given central temperature, our bare cores are denser in the center and cooler in the outer parts compared to their equivalent embedded cores. In bare cores, the carbon ignites slightly earlier with a higher energy generation and smaller neutrino loss rates. The change in the central carbon burning in turn starts a ripple effect by modifying the timing, location, and the extent of the next shell and core burning episodes. Higher mass cores are most sensitive to these changes since carbon burns radiatively in the central core with a very short lifetime as compared to lower mass cores with long lasting convective burning. These are not only the consequences of the artificially constant mass and boundary pressure in our models, but are also influenced by the intrinsically different outer structure. There is no helium burning shell in bare CO-cores, and the strength of the outermost carbon burning shell is often very different than in the embedded counterparts. In a sense, the final core structure in \emph{bare} CO-cores changes more slowly with increasing mass compared to embedded cores, and hence forms a much broader \cp\ peak. This effect is more pronounced for full stars, and it is only mildly present with He stars.

Though the calculated and interpolated values of \cp\ are not in full agreement with one another, they do follow the same overall trends. However, these simple first iteration CO-core models should not be used to infer the final structure of a given stellar model, and instead should only be applied to populations of massive stars to infer the underlying trends in the properties of their final demise.

\begin{figure*}
    \centering
    \hfill\includegraphics[scale=0.70,trim={0.5cm 0.1cm 0.5cm 0.1cm},clip]{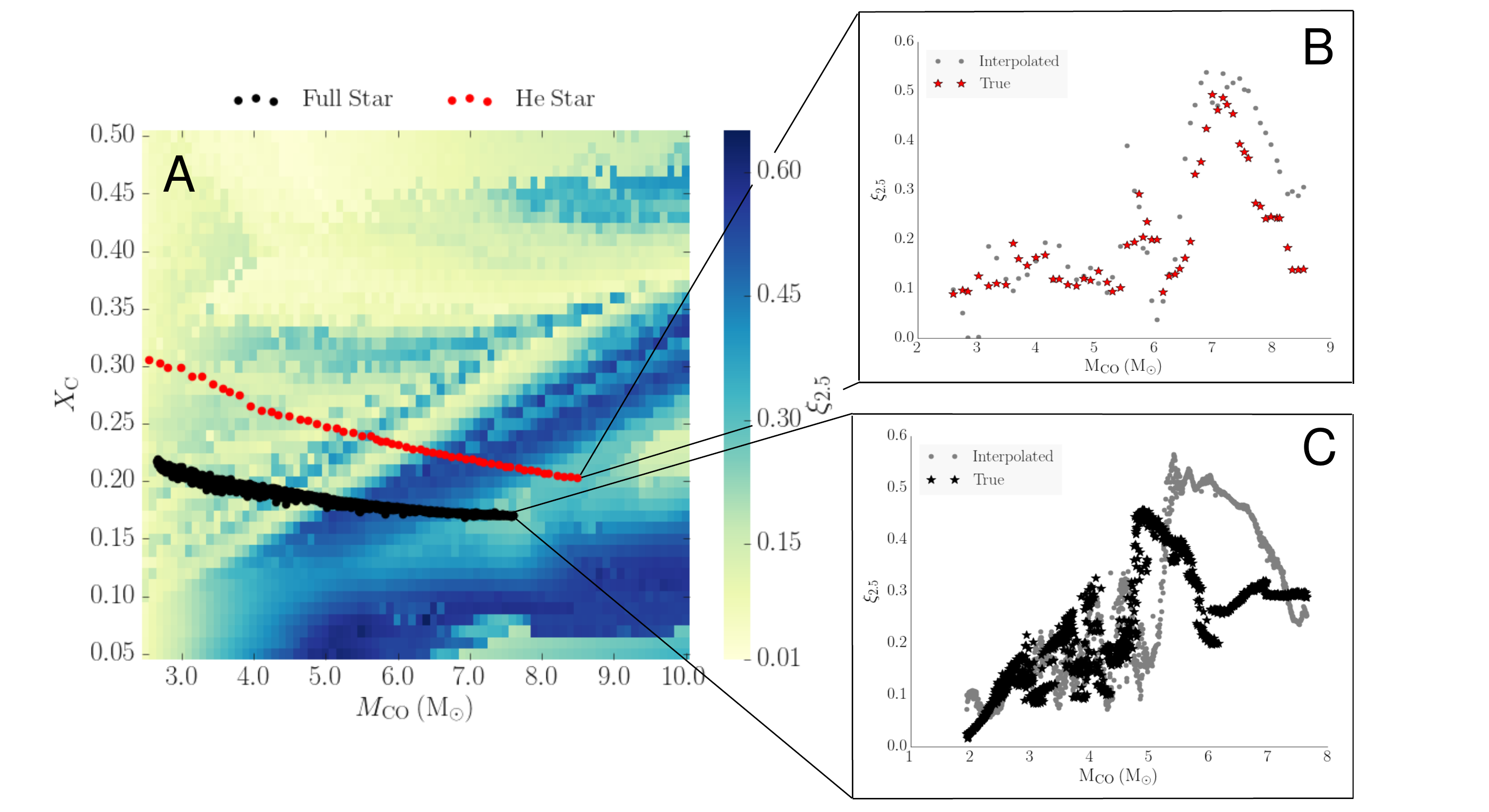}\hspace*\fill
    \caption{Panel A shows the core parameters from full stars (black) and He stars (red) superimposed on the \cp~distribution (same as in \Fig{3par}). For a given set of single star models, the CO-cores start out with more oxygen with increasing mass. Panels B and C compare the values of \cp~ for the full and He star models calculated directly from the stellar presupernova core properties and those interpolated from our tables. While the height and width of the peak in \cp\ at higher core masses varies between the true and interpolated values, the bulk trends remain the same.}
    \lFig{full_comp}
\end{figure*}


\section{Implementation}
\lSect{impl}

The application of our approach requires knowledge of the CO-core mass and its initial carbon mass fraction for each model. Binary population synthesis calculations that employ stellar evolution codes often reach the central carbon burning phase, and both CO-core properties can be easily extracted. In these cases our tables can be used directly to infer the core structure at the onset of collapse, and below we provide a sample application to recent results on Type-IIb supernova progenitors. Our results cannot be directly applied to ``rapid'' BPS calculations, in which only the CO-core mass is tracked, but not its starting composition. At the end of this section we discuss possible ways to approximate $X_{\rm C}$ in such scenarios.

\subsection{Sample Application: progenitors of Type-IIb supernovae}
\lSect{sra}

Given the simplistic nature of existing prescriptions used to determine the final fates of stars, we expect the outcomes of many BPS studies to be affected by our results. As a test case, we apply our calculations to a recent study on Type-IIb supernova progenitors by \citet{Sra19}. Using \MESA\ they have simulated a large population of single and binary models and find, among other things, that the binary channel dominates at low metallicity and may fully account for the observed rate of Type-IIb events. A full description of their parameter space is listed in their Table 1, but in summary, binary models were explored within primary ZAMS masses of $1.0 < \log\ M_1 < 1.4$ \Msun\ , mass ratios of $0.225 < q < 0.975$, mass transfer efficiencies of $0.01 < \epsilon < 1.0$, initial orbital periods of $1.0 < \log\ P < 3.8$ days, and metallicities of \Zsun\ and $\Zsun/4$. For solar metallicity models, the span in initial period is reduced to $2.5 < \log\ P < 3.8$ days. To be considered a binary system in their study, the primary star must lose at least 1\% of its initial mass to Roche lobe overflow. All primary stars which reached central carbon depletion while retaining hydrogen envelopes with masses between 0.01 and 1 \Msun\ were designated as Type-IIb progenitors. 

To produce a Type-IIb supernova, the progenitor star not only needs to reach core-collapse with the right envelope properties, but the collapse ultimately needs to produce an explosion. Although the upper bound on $M_1$ ($\sim$25 \Msun) in their grid excludes most stars that would be difficult to explode with neutrinos, it does include many stars that are likely to implode, and thus, would not produce a Type-IIb despite having the right envelope mass. We test this by recomputing a sample subset of their population at a fixed mass ratio ($q=0.925$) and mass transfer efficiency ($\epsilon=0.1$), using the same version of \MESA\ (9575). Since the lightest CO-cores produce explosions over a wide range of initial compositions, we only explore primary stars more massive than about 14.5 \Msun\ (roughly corresponding to $M_{\rm CO}\sim3$ \Msun). Otherwise we probe their entire parameter space in mass and orbital period using increments of 0.04 and 0.05 dex respectively, about double the increments used in their paper, which they note should not affect the results. 

All models were terminated at either carbon ignition or when the envelope mass dropped beneath 0.01 \Msun. The former is slightly before the evolutionary cutoff used in \citet{Sra19} (carbon depletion), but due to the negligible mass lost during carbon burning, our results are nearly identical to theirs. Following the \citet{Sra19} criteria, the models which retained a hydrogen envelope of 0.01 - 1 \Msun\ at carbon ignition and lost at least 1\% of the primary star`s initial mass to Roche lobe overflow were designated as binary Type-IIb progenitors. We then took the core properties, $M_\mathrm{CO}$ and $X_\mathrm{C}$, at carbon ignition and rounded to the nearest discrete point in our grid of final fates from \Fig{ertl}, adopting the outcome of that point. As in \Sect{comp}, we take $M_\mathrm{CO}$ to be the average of the 1\% and 25\% threshold masses.

\begin{figure}
    \centering
    \includegraphics[scale=0.48,trim={0.7cm 0.0cm 0.7cm 0.7cm},clip]{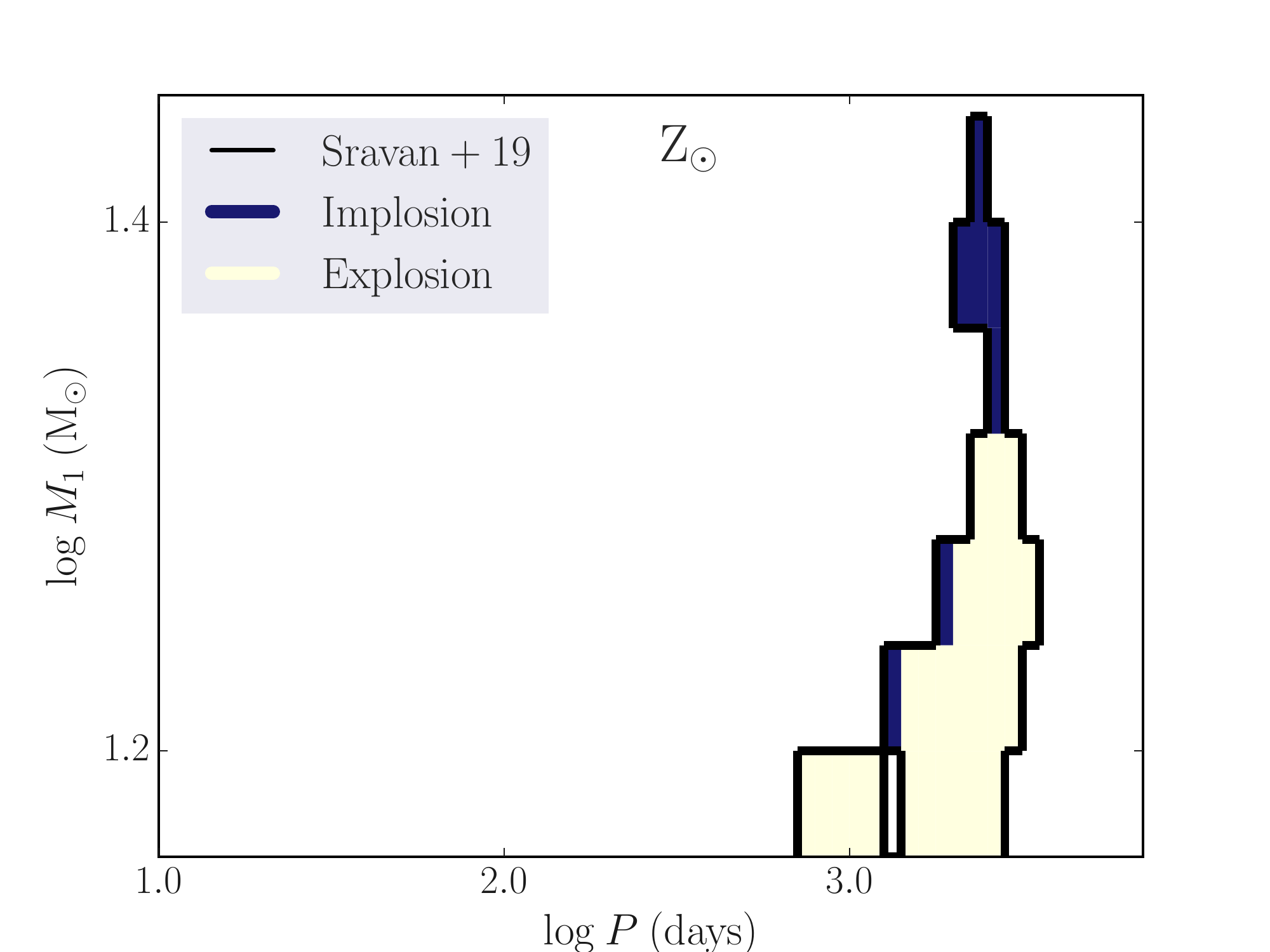}
    \includegraphics[scale=0.48,trim={0.7cm 0.0cm 0.7cm 0.7cm},clip]{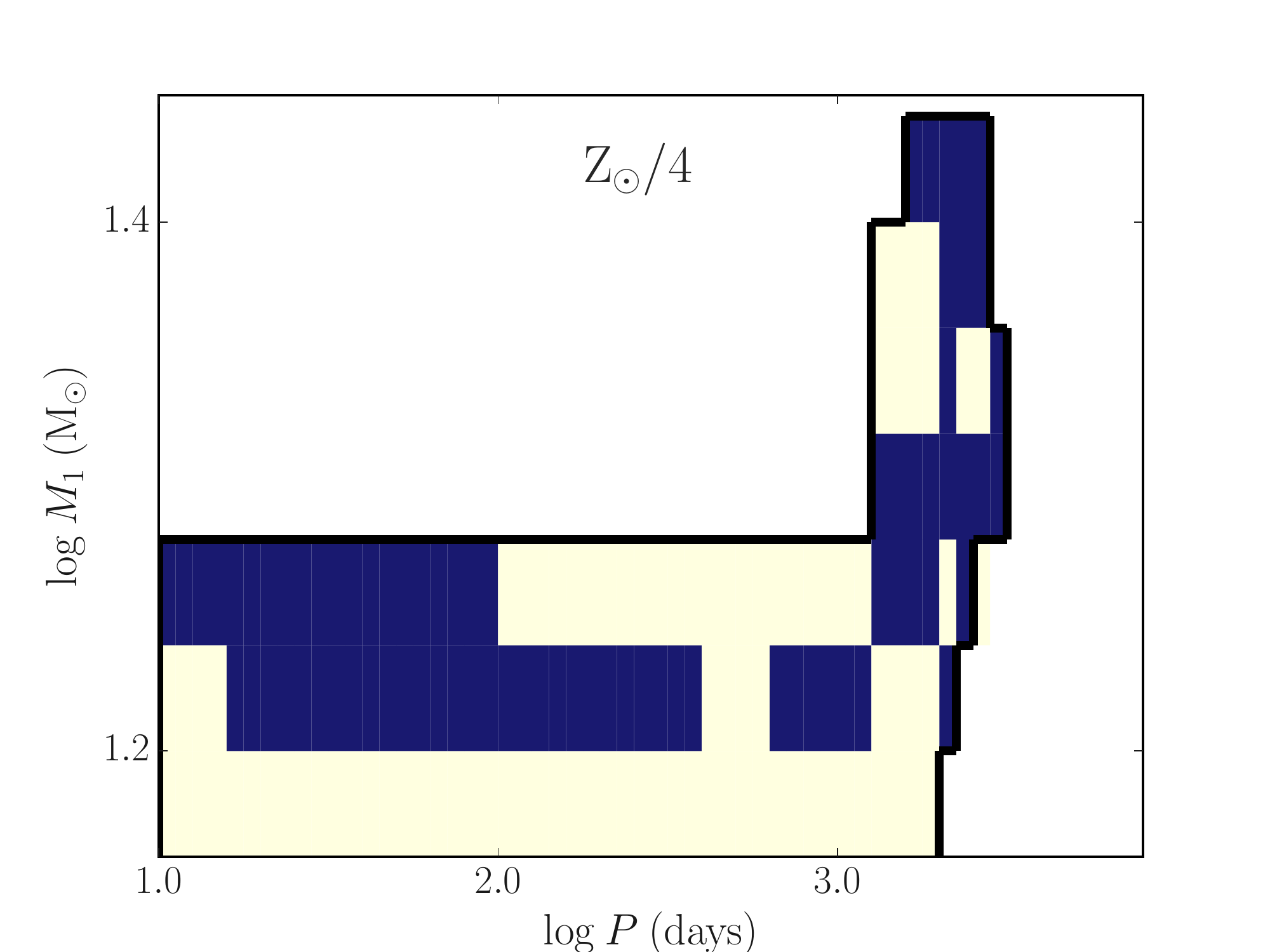}
    \caption{The regions of the primary mass - binary orbital period parameter space which produce Type-IIb supernova progenitors using the criteria in \citet{Sra19} (outlined in black) and our core models (light bars explode, dark bars implode). The top panel shows the results for solar metallicity models and the bottom shows \Zsun/4 models. Given the presupernova structures inferred from our CO-core models, and the application of a sample explosion criterion on them, we find that significant fraction of stars that collapse with the right hydrogen envelope mass may not be likely to produce a Type-IIb supernova explosion.}
    \lFig{sra}
\end{figure}

\Fig{sra} shows the results from both the \citet{Sra19} criteria and our CO-core models. The areas outlined in black highlight the region of the parameter space which produce Type-IIb supernovae according to the \citet{Sra19} criteria. At solar metallicity (top panel), progenitors with low mass envelope form exclusively in wide systems, while at lower metallicity (bottom panel), they can form in shorter period systems due to weaker envelope expansion and winds. These results fully match their results (see their Figures 4 and 5). However, we find that only a subset of these models produce supernova explosions when we consider their presupernova core structures based on our CO-core models and apply the same sample explosion criterion from \Sect{ertl}. All light colored regions enclosed in the outlined areas explode and all darker regions implode. The most massive models, at both metallicities, have radiative central carbon burning and result in implosion as they lie beneath the diagonal line in the $X_{\rm C}-M_{\rm CO}$ plane. Lighter models with convective central carbon burning generally explode, but at low metallicity, many of the shorter period stars do not explode as they cross the region just above the diagonal line where the final outcomes are uncertain (see \Sect{expl}). By number, we find roughly a fifth of the models in our small grid implode at \Zsun, and roughly half at \Zsun/4.

Although one needs to examine other combinations of $q$ and $\epsilon$, and apply proper initial mass function weighting, these results already pose potential challenges to some of the findings by \citet{Sra19}. For instance, the observed rate of Type-IIb events is unlikely to be fully accounted for by low metallicity systems, even with low mass transfer efficiency because many of these primary stars implode instead of explode. The relative contributions from single and binary channels, and its dependence on metallicity, could be be significantly different as well.

\subsection{Only having $M_{\rm CO}$}
\lSect{Xc_est}

The rapid BPS codes \citep[e.g.,][]{Gia18a,Bre19} approximate stellar evolution through the fits by \citet{Hur00,Hur02}, which only compute the CO-core mass instead of both mass and composition. The relation between initial carbon mass fraction and $M_{\rm CO}$ is sensitive to the input physics operating during the prior evolution (e.g., reaction rates, mixing, rotation), and it cannot be reliably approximated, especially for populations spanning different metallicities. The ideal solutions are to incorporate a dedicated prescription in the BPS code to track $X_{\rm C}$ or to map it by employing stellar evolution calculations for a subset of the population.

A very basic approximation may be made by utilizing the existing single star models and by assuming that the relation between $X_{\rm C}$ and $M_{\rm CO}$ changes only depending on whether the He-core in each star was embedded or exposed during the evolution. For example, the solar metallicity single star models from \citet{Suk18} and helium stars evolved with mass loss from \citet{Woo19} give (see \Fig{full_comp}A)
\begin{equation}
    X_{\rm C} = 0.20/M_{\rm CO} + 0.15
\end{equation}
for single stars and
\begin{equation}
    X_{\rm C} = -0.084 \times \ln\ (M_{\rm CO}) + 0.4
\end{equation}
for mass losing, bare, helium stars.

The initial carbon mass fractions for stars that have lost their envelopes through binary interaction could be estimated by using the relation based on mass losing helium star models, and the relation based on full star models could be used for all other stars in which the He-cores were always embedded. However, such an approximation is highly simplistic, and completely ignores the fact that even for single stars the relation between $X_{\rm C}$ and $M_{\rm CO}$ can appreciably change due to uncertain input physics and variations in metallicity.

\subsection{Beyond Final Fates}

While in this study we focus on the final fates, our results can be utilized to infer other properties. For example, the presupernova structures can be used in conjunction with some explosion criterion to approximate the compact remnant mass. When a star explodes based on its embedded CO-core, the baryonic mass of the resulting neutron star could be approximated by $M_4$. When the embedded CO-core is deemed to implode, the resulting black hole mass could be bracketed as the helium core mass at the evolutionary cutoff, as a minimum, plus some uncertain but small fraction of the surviving envelope (if the star retains some envelope) as a maximum. These results also could be used to infer the nucleosynthesis, however, this requires a coupling to an explosion modelling, and will be explored in a forthcoming study. 

\section{Conclusions}
\lSect{concl}

Knowing the final fate of the star is critical in any population study involving massive stars. Only exploding stars produce bright supernovae, substantially contribute to nucleosynthetic yields, and dissolve binary systems. The final fate of the massive star, whether it explodes or implodes, is dependent on its core structure right before the collapse. However, the majority of current population synthesis calculations follow the evolution of stars only until central carbon burning (\Fig{comp}), and without knowing the final presupernova structure, they resort to simple prescriptions to infer their final fates.

In this study we explore one potential solution to this problem by treating the CO-cores independently from the rest of the star. Through an extensive suite of bare CO-core models, each evolved from central carbon ignition until the onset of core-collapse, for the first time we map final presupernova core structures to a diverse set of CO-core initial conditions. These models reasonably mimic the embedded CO-cores of massive stars, and can be utilized in single and/or binary population synthesis calculations to infer the final fates and the remnant properties based on the presupernova core structure. We summarize the main results as follows:

\begin{enumerate}

\item Our main grid consists of 3496 models computed with the implicit hydrodynamics code \KEPLER. For simplicity the mass of the CO-core was assumed to be constant and all models start from a uniform mixture of carbon and oxygen only. The range of initial compositions ($0.5>X_{\rm C}>0$) encompasses nearly all combinations possible in nature, while the mass was confined between 2.5 and 10 \Msun. All lighter CO-cores that reach iron-core collapse will likely produce a supernova explosion no matter the starting composition, while all heavier cores will implode until the onset of pair-instability effects. Our grid captures the mass range where the final core structure, and therefore the final fate of the core, varies the most depending on its initial composition. 

\item We have explored the presupernova core structures by focusing on three simple parameters -- $M_{\rm Fe}$, $M_4$, and $\cp$. In the $X_{\rm C}-M_{\rm CO}$ plane all three parameters exhibit similar trends reflecting their close correlation (\Fig{3par}). They show that the final core structure can vary substantially for cores that start from identical mass but different compositions or from identical composition but different masses. The plane is broadly separated into two key regions separated by a diagonal line stretching from roughly $M_{\rm CO}\approx2.5$ \Msun\ and $X_{\rm C}\approx0.10$ to about $M_{\rm CO}\approx10$ \Msun\ and $X_{\rm C}\approx0.4$, above (below) which carbon burns convectively (radiatively) in the center. Most of the cores with convective central carbon burning die compact with smaller $M_{\rm Fe}$, $M_4$, and $\cp$, but the opposite is true for those with radiative central carbon burning. However, in agreement with prior studies, the final structure varies highly non-monotonically, with compact final cores forming beneath the diagonal line and extended final cores above it.

\item Applying a sample explosion criterion, we find that the final fates of these cores closely correlate with the presupernova structure (\Fig{ertl}). Most stars which form smaller $M_{\rm Fe}$, $M_4$, and $\cp$ end up exploding, and vice versa. Though the outcomes are strongly delineated by the central carbon burning transition, there are substantial non-monotonic variations. Based on the distribution of three parameters and the associated outcomes from the adopted explosion criterion, we find the values that separate most of the explosions from most of the implosions to be $M_{\rm Fe}=1.54$ \Msun, $M_4=1.66$ \Msun, and $\cp=0.18$ (\Fig{ertl} and \Tab{ertl_tab}).

\item We test the CO-core model results by recreating a subset of our grid through an independent open source stellar evolution code \MESA. Though there are slight differences due to input setup, there is good overall agreement (\Fig{mesa1} and \Fig{mesa2}). We have also compared our calculations with red supergiant and helium star models embedding equivalent CO-cores (\Fig{full_comp}). While the variations are in good agreement, especially for the helium stars and the lighter hydrogenic stars, there are substantial differences that point to the limitations of our approach. Our models are not perfect substitutes for the embedded CO-cores of full stars, and they should only be applied to infer bulk trends in populations rather than individual models.

\item Our results are easy to use. A table listing $M_{\rm Fe}$, $M_4$, $\mu_4$ and $\cp$ values, as well as the full presupernova structure and composition for each model are available for download \footnotemark[2]. A sample explosion criterion is discussed in \Sect{ertl}, but we leave it up to the reader to explore other options. The initial conditions of the CO-cores, $X_{\rm C}$ and $M_{\rm CO}$, can be easily extracted in population synthesis calculations that employ active or passive stellar evolution, and our results can be utilized as fast lookup tables. In \Sect{sra} we demonstrate a sample application on a recent study on Type-IIb supernova progenitors. Though our results cannot be applied directly to rapid BPS calculations, in \Sect{Xc_est} we provide fitting functions for a simple estimate of $X_{\rm C}$ based on the CO-core mass.

\end{enumerate}

Our future work will expand on a number of fronts. A followup study will be focused on the late stage evolution of CO-cores, and will explore ways to model them to even more closely mimic embedded counterparts. We are also planning to develop a prescription for BPS codes to estimate the initial carbon mass fraction, $X_{\rm C}$. The explodability of these cores will be explored in the context of the neutrino-driven mechanism and nucleosynthesis as well.

\section{Acknowledgements}
The authors would like to thank C. S. Kochanek and S. E. Woosley for many helpful comments. We also thank S. E. Woosley for sharing the data on his helium star models, and N. Sravan for providing the details of her \MESA\ models. Numerical calculations using the \KEPLER\ and \MESA\ codes were performed on the \texttt{RUBY} cluster at the Ohio Supercomputer Center \citep{osc}. Support for this work comes from the National Science Foundation through grant AST-1515927 awarded to K. Z. Stanek and C. S. Kochanek. TS was supported by NASA through the NASA Hubble Fellowship grant \#60065868 awarded by the Space Telescope Science Institute, which is operated by the Association of Universities for Research in Astronomy, Inc., for NASA, under contract NAS5-26555. 



\bsp	
\label{lastpage}

\begin{thebibliography}{99}

\bibitem[Arnett(1972a)]{Arn72a} 
Arnett, W.~D.\ 1972, \apj, 173, 393

\bibitem[Arnett(1972b)]{Arn72b} 
Arnett, W.~D.\ 1972, \apj, 176, 681

\bibitem[Barkat \& Marom(1990)]{Bar90}
Barkat, Z., \& Marom, A.\ 1990, Supernovae, Jerusalem Winter School for Theoretical Physics, ed. J.C. Wheeler, T. Piran, S. Weinberg; (World Scientific Publishing Co.), 95 

\bibitem[Belczynski et al.(2002a)]{Bel02a}
Belczynski, K., Kalogera, V., \& Bulik, T.\ 2002, \apj, 572, 407

\bibitem[Belczynski et al.(2002b)]{Bel02b}
Belczynski, K., Bulik, T., \& Klu{\'z}niak, W.\ 2002, \apjl, 567, L63

\bibitem[Belczynski et al.(2016)]{Bel16} Belczynski, K., Repetto, S., Holz, D.~E., et al.\ 2016, \apj, 819, 108

\bibitem[Bennett et al.(2012)]{Ben12}
Bennett, M.~E., Hirschi, R., Pignatari, M., et al.\ 2012, \mnras, 420, 3047

\bibitem[Breivik et al.(2019)]{Bre19}
Breivik, K., Coughlin, S.~C., Zevin, M., et al.\ 2019, arXiv e-prints, arXiv:1911.00903

\bibitem[Buchmann(1996)]{Buc96}
Buchmann, L.\ 1996, \apjl, 468, L127

\bibitem[Burrows et al.(1995)]{Bur95}
Burrows, A., Hayes, J., \& Fryxell, B.~A.\ 1995, \apj, 450, 830.

\bibitem[Burrows et al.(2020)]{Bur20} 
Burrows, A., Radice, D., Vartanyan, D., et al.\ 2020, \mnras, 491, 2715

\bibitem[Chrimes et al.(2019)]{Chr19}
Chrimes, A.~A., Stanway, E.~R., \& Eldridge, J.~J.\ 2019, \mnras, 2830

\bibitem[De Donder \& Vanbeveren(2004)]{DeD04}
De Donder, E., \& Vanbeveren, D.\ 2004, \nar, 48, 861

\bibitem[de Mink et al.(2014)]{deM14}
de Mink, S.~E., Sana, H., Langer, N., et al.\ 2014, \apj, 782, 7

\bibitem[Ebinger et al.(2019)]{Ebi19}
Ebinger, K., Curtis, S., Fr{\"o}hlich, C., et al.\ 2019, \apj, 870, 1

\bibitem[Eldridge et al.(2008)]{Eld08}
Eldridge, J.~J., Izzard, R.~G., \& Tout, C.~A.\ 2008, \mnras, 384, 1109

\bibitem[Eldridge \& Stanway(2016)]{Eld16}
Eldridge, J.~J., \& Stanway, E.~R.\ 2016, \mnras, 462, 3302

\bibitem[Eldridge et al.(2017)]{Eld17}
Eldridge, J.~J., Stanway, E.~R., Xiao, L., et al.\ 2017, \pasa, 34, e058

\bibitem[Ertl et al.(2016)]{Ert16}
Ertl, T., Janka, H.-T., Woosley, S.~E., et al.\ 2016a, \apj, 818, 124

\bibitem[Ertl et al.(2020)]{Ert19}
Ertl, T., Woosley, S.~E., Sukhbold, T., et al.\ 2020, \apj, 890, 51

\bibitem[Farmer et al.(2016)]{Far16}
Farmer, R., Fields, C.~E., Petermann, I., et al.\ 2016, \apjs, 227, 22

\bibitem[Fraser et al.(2013)]{Fra13} 
Fraser, M., Magee, M., Kotak, R., et al.\ 2013, \apjl, 779, L8

\bibitem[Fryer et al.(2012)]{Fry12} 
Fryer, C.~L., Belczynski, K., Wiktorowicz, G., et al.\ 2012, \apj, 749, 91 

\bibitem[Fuller(2017)]{Ful17}
Fuller, J.\ 2017, \mnras, 470, 1642

\bibitem[Giacobbo \& Mapelli(2018)]{Gia18a}
Giacobbo, N., \& Mapelli, M.\ 2018, \mnras, 480, 2011

\bibitem[Giacobbo et al.(2018)]{Gia18b}
Giacobbo, N., Mapelli, M., \& Spera, M.\ 2018, \mnras, 474, 2959

\bibitem[Heger et al.(2001)]{Heg01} 
Heger, A., Woosley, S.~E., Mart{\'\i}nez-Pinedo, G., et al.\ 2001, \apj, 560, 307

\bibitem[Heger et al.(2003)]{Heg03}
Heger, A., Fryer, C.~L., Woosley, S.~E., et al.\ 2003, \apj, 591, 288

\bibitem[Horiuchi et al.(2014)]{Hor14} 
Horiuchi, S., Nakamura, K., Takiwaki, T., et al.\ 2014, \mnras, 445, L99

\bibitem[Hurley et al.(2000)]{Hur00}
Hurley, J.~R., Pols, O.~R., \& Tout, C.~A.\ 2000, \mnras, 315, 543

\bibitem[Hurley et al.(2002)]{Hur02}
Hurley, J.~R., Tout, C.~A., \& Pols, O.~R.\ 2002, \mnras, 329, 897

\bibitem[Izzard et al.(2004)]{Izz04} 
Izzard, R.~G., Tout, C.~A., Karakas, A.~I., et al.\ 2004, \mnras, 350, 407

\bibitem[Izzard et al.(2006)]{Izz06} 
Izzard, R.~G., Dray, L.~M., Karakas, A.~I., et al.\ 2006, \aap, 460, 565

\bibitem[Izzard et al.(2009)]{Izz09} 
Izzard, R.~G., Glebbeek, E., Stancliffe, R.~J., et al.\ 2009, \aap, 508, 1359

\bibitem[Janka et al.(2016)]{Jan16}
Janka, H.-T., Melson, T., \& Summa, A.\ 2016, Annual Review of Nuclear and Particle Science, 66, 341

\bibitem[Johnson et al.(2018)]{Joh18} 
Johnson, S.~A., Kochanek, C.~S., \& Adams, S.~M.\ 2018, \mnras, 480, 1696

\bibitem[Jones et al.(2013)]{Jon13}
Jones, S., Hirschi, R., Nomoto, K., et al.\ 2013, \apj, 772, 150

\bibitem[Jones et al.(2016)]{Jon16}
Jones, S., R{\"o}pke, F.~K., Pakmor, R., et al.\ 2016, \aap, 593, A72

\bibitem[Kochanek et al.(2017)]{Koc17} 
Kochanek, C.~S., Fraser, M., Adams, S.~M., et al.\ 2017, \mnras, 467, 3347

\bibitem[Kruckow et al.(2018)]{Kru18}
Kruckow, M.~U., Tauris, T.~M., Langer, N., et al.\ 2018, \mnras, 481, 1908

\bibitem[Limongi \& Chieffi(2018)]{Lim18} 
Limongi, M., \& Chieffi, A.\ 2018, \apjs, 237, 13

\bibitem[Mabanta et al.(2019)]{Mab19} Mabanta, Q.~A., Murphy, J.~W., \& Dolence, J.~C.\ 2019, \apj, 887, 43

\bibitem[Mauerhan et al.(2013)]{Mau13} 
Mauerhan, J.~C., Smith, N., Filippenko, A.~V., et al.\ 2013, \mnras, 430, 1801

\bibitem[Mazurek(1982)]{Maz82}
Mazurek, T.~J.\ 1982, \apjl, 259, L13

\bibitem[Meakin \& Arnett(2006)]{Mea06} 
Meakin, C.~A., \& Arnett, D.\ 2006, \apjl, 637, L53

\bibitem[Meakin \& Arnett(2007)]{Mea07} 
Meakin, C.~A., \& Arnett, D.\ 2007, \apj, 667, 448

\bibitem[Melson et al.(2015)]{Mel15} 
Melson, T., Janka, H.-T., \& Marek, A.\ 2015, \apjl, 801, L24

\bibitem[Menon \& Heger(2017)]{Men17}
Menon, A., \& Heger, A.\ 2017, \mnras, 469, 4649 

\bibitem[M{\"u}ller et al.(2016)]{Mue16}
M{\"u}ller, B., Heger, A., Liptai, D., et al.\ 2016, \mnras, 460, 742

\bibitem[Nomoto \& Hashimoto(1988)]{Nom88} 
Nomoto, K., \& Hashimoto, M.\ 1988, \physrep, 163, 13

\bibitem[O'Connor \& Ott(2011)]{OCo11}
 O'Connor, E., \& Ott, C.~D.\ 2011, \apj, 730, 70

\bibitem[Ohio Supercomputer Center (1987)]{osc}
Ohio\ Supercomputer\ Center\ 1987, Ohio\ Supercomputer\ Center, http://osc.edu/ark:/19495/f5s1ph73

\bibitem[Pastorello et al.(2007)]{Pas07} 
Pastorello, A., Smartt, S.~J., Mattila, S., et al.\ 2007, \nat, 447, 829

\bibitem[Paxton et al.(2010)]{Pax10} 
Paxton, B., Bildsten, L., Dotter, A., et al.\ 2010, MESA: Modules for Experiments in Stellar Astrophysics, ascl:1010.083


\bibitem[Paxton et al.(2019)]{Pax19} 
Paxton, B., Smolec, R., Schwab, J., et al.\ 2019, \apjs, 243, 10

\bibitem[Pejcha \& Thompson(2015)]{Pej15}
 Pejcha, O., \& Thompson, T.~A.\ 2015, \apj, 801, 90 
 
 \bibitem[Podsiadlowski et al.(2007)]{Pod07} 
 Podsiadlowski, P., Morris, T.~S., \& Ivanova, N.\ 2007, Supernova 1987A: 20 Years After: Supernovae and Gamma-ray Bursters, 125

\bibitem[Renzo et al.(2017)]{Ren17}
Renzo, M., Ott, C.~D., Shore, S.~N., \& de Mink, S.~E.\ 2017, \aap, 603, A118

\bibitem[Renzo et al.(2019)]{Ren19}
Renzo, M., Zapartas, E., de Mink, S.~E., et al.\ 2019, \aap, 624, A66

\bibitem[Sana et al.(2012)]{San12} 
Sana, H., de Mink, S.~E., de Koter, A., et al.\ 2012, Science, 337, 444

\bibitem[Sana et al.(2013)]{San13} 
Sana, H., de Koter, A., de Mink, S.~E., et al.\ 2013, \aap, 550, A107

\bibitem[Shiode, \& Quataert(2014)]{Shi14}
Shiode, J.~H., \& Quataert, E.\ 2014, \apj, 780, 96

\bibitem[Siess et al.(2013)]{Sie13}
Siess, L., Izzard, R.~G., Davis, P.~J., et al.\ 2013, \aap, 550, A100

\bibitem[Spera et al.(2015)]{Spe15}
Spera, M., Mapelli, M., \& Bressan, A.\ 2015, \mnras, 451, 4086

\bibitem[Spera \& Mapelli(2017)]{Spe17}
Spera, M., \& Mapelli, M.\ 2017, \mnras, 470, 4739

\bibitem[Spera et al.(2019)]{Spe19}
Spera, M., Mapelli, M., Giacobbo, N., et al.\ 2019, \mnras, 485, 889

\bibitem[Sukhbold \& Woosley(2014)]{Suk14}
 Sukhbold, T., \& Woosley, S.~E.\ 2014, \apj, 783, 10

\bibitem[Sukhbold et al.(2016)]{Suk16}
Sukhbold, T., Ertl, T., Woosley, S.~E., et al.\ 2016, \apj, 821, 38

\bibitem[Sukhbold et al.(2018)]{Suk18}
Sukhbold, T., Woosley, S.~E., \& Heger, A.\ 2018, \apj, 860, 93

\bibitem[Sukhbold \& Adams(2020)]{Suk20} 
Sukhbold, T., \& Adams, S.\ 2020, \mnras, 492, 2578

\bibitem[Sravan et al.(2019)]{Sra19}
Sravan, N., Marchant, P., \& Kalogera, V.\ 2019, \apj, 885, 130

\bibitem[Tur et al.(2007)]{Tur07}
Tur, C., Heger, A., \& Austin, S.~M.\ 2007, \apj, 671, 821

\bibitem[Ugliano et al.(2012)]{Ugl12}
 Ugliano, M., Janka, H.-T., Marek, A., \& Arcones, A.\ 2012, \apj, 757, 69

\bibitem[Vigna-G{\'o}mez et al.(2018)]{Vig18}
Vigna-G{\'o}mez, A., Neijssel, C.~J., Stevenson, S., et al.\ 2018, \mnras, 481, 4009

\bibitem[Weaver et al.(1978)]{Wea78}
Weaver, T.~A., Zimmerman, G.~B., \& Woosley, S.~E.\ 1978, \apj, 225,1021

\bibitem[Woosley(2017)]{Woo17} 
Woosley, S.~E.\ 2017, \apj, 836, 244

\bibitem[Woosley(2019)]{Woo19} 
Woosley, S.~E.\ 2019, \apj, 878, 49

\bibitem[Woosley \& Heger(2015)]{Woo15}
Woosley, S.~E., \& Heger, A.\ 2015, \apj, 810, 34

\bibitem[Woosley et al.(2002)]{Woo02}
Woosley, S.~E., Heger, A., \& Weaver, T.~A.\ 2002, Reviews of Modern Physics, 74, 1015

\bibitem[Woosley et al.(2020)]{Woo20}
Woosley, S., Sukhbold, T., \& Janka, H.-T.\ 2020, arXiv e-prints, arXiv:2001.10492

\bibitem[Woosley \& Weaver(1995)]{Woo95} 
Woosley, S.~E., \& Weaver, T.~A.\ 1995, \apjs, 101, 181

\bibitem[Yoon et al.(2010)]{Yoo10} 
Yoon, S.-C., Woosley, S.~E., \& Langer, N.\ 2010, \apj, 725, 940

\bibitem[Zapartas et al.(2019)]{Zap19}
Zapartas, E., de Mink, S.~E., Justham, S., et al.\ 2019, \aap, 631, A5
\end{thebibliography}
\end{document}